\documentclass{article}

\usepackage{arxiv}

\usepackage[utf8]{inputenc} 
\usepackage[T1]{fontenc}    
\usepackage{hyperref}       
\usepackage{url}            
\usepackage{booktabs}       
\usepackage{amsfonts}       
\usepackage{nicefrac}       
\usepackage{microtype}      
\usepackage{lipsum}		
\usepackage{graphicx}
\usepackage[numbers,sort&compress]{natbib}
\usepackage{amsmath,amssymb}
\usepackage{booktabs}
\usepackage{placeins}
\usepackage{subcaption}

\title{Supervised Source Separation of Electric Vehicle Interior Noise Using a Hybrid Transformer Architecture}

\date{} 					

\author{Philipp Diel$^{1,}$\thanks{Corresponding author: \texttt{diel@tme.rwth-aachen.de}}
  \And
  Zhuo Cao$^{2}$
  \And
  Arne Hildenbrand$^{3}$
  \And
  Marco Günther$^{1}$
  \AND
  Stefan Pischinger$^{1}$
  \And
  Hanno Scharr$^{2}$
  \And
  Steffen Marburg$^{3}$
  \AND
  \\[0pt]
  \small $^{1}$Chair of Thermodynamics of Mobile Energy Conversion Systems, RWTH Aachen University, Aachen, Germany \\
  \small $^{2}$ IAS-8, Forschungszentrum Jülich, Jülich, Germany \\
  \small $^{3}$Chair of Vibroacoustics of Vehicles and Machines, Technical University of Munich, Garching, Germany
}

\renewcommand{\headeright}{}
\renewcommand{\shorttitle}{Supervised Source Separation of Electric Vehicle Interior Noise Using a Hybrid Transformer Architecture}

\hypersetup{
pdftitle={},
pdfsubject={},
pdfauthor={},
pdfkeywords={},
}

\begin{document}
\maketitle


\begin{abstract}
The interior sound of an electric vehicle is a superposition of several distinct noise contributions -- wind noise, tire noise, powertrain (electric drive unit and inverter) noise, and auxiliary-drive noise -- whose relative balance governs both the overall loudness and the perceived sound quality in the vehicle interior. Decomposing a recorded vehicle interior signal into these source contributions is a prerequisite for targeted acoustic engineering, but requires effortful measurement campaigns as well as expertise in the field of transfer path analysis. This work formulates the decomposition as a supervised, binaural source-separation problem with four target source contributions and adapts a hybrid time-spectrogram transformer architecture, originally developed for music demixing, to the domain of sound source separation for technical systems.
To disambiguate source contributions, conditioning is introduced in the form of operating-state information, including electric drive speed, electric drive load, vehicle speed, and auxiliary-drive speed. Since isolated real-world reference stems are costly to obtain, the model is trained on a physics-based parametric data generator that synthesizes labeled component sounds from randomized virtual-vehicle hardware configurations and driving scenarios, so that the ground-truth contribution of each source is known by construction. Validation on both synthetic and measured vehicle interior sounds probes the achieved separation performance and underscores the feasibility of applying the proposed method for acoustic contribution analysis in technical systems.

\end{abstract}



\maketitle

\section{Introduction}
\label{sec:introduction}
In vehicles with an internal combustion engine, the broadband and low-frequency engine noise acts as a masking noise that conceals many secondary sources. In battery electric vehicles (BEVs) and fuel cell electric vehicles (FCEVs), this masking noise is absent, so that the remaining contributions -- wind noise, tire noise, noise of the electric drive unit and its inverter, and noise of auxiliary components -- determine the vehicle interior sound.
The engineering of the interior sound of an EV is therefore concerned with the balance of these individual noise contributions. Thus, separation of the measured interior noise into its individual contributions is a useful tool for the acoustic development process, as it allows the perceived annoyance to be attributed to specific sources, the acoustic masking relations to be quantified, and the effect of design changes on the individual contributions to be assessed. This work investigates whether such a separation can be performed automatically by means of a learned separation model and reports the attainable separation quality in the given setup.

Beyond this, we ask whether a model trained exclusively on physics-based synthetic data can be applied to measured vehicle interior noise, that is, whether the simulation-to-reality gap can be bridged by domain randomization of the virtual vehicle hardware. Both questions are posed with a practical decision in mind: whether the acquisition of measured training data with known source contributions -- which requires transfer path analysis campaigns and is therefore costly -- is justified by the separation quality that a learned model can be expected to reach.

Answering these questions requires bringing together two bodies of work that are usually treated separately: the acoustics of electric vehicle interior noise and its established, test-based contribution analysis on one side, and data-based audio source separation, developed mainly for music demixing and speech separation, on the other. The sources considered here are technical noise contributions with distinct physical excitation mechanisms that are permanently active, differ by orders of magnitude in level, and for which isolated reference recordings are not readily available. The following sections review both fields in turn and derive the resulting design choices.

\subsection{Interior Noise of Electric Vehicles}

Noise generation due to the tire-road contact during driving is a complex phenomenon that has garnered high interest in the scientific community \citep{LI2018tirereview}. Tire noise is transmitted to the vehicle cabin both via structural vibration and airborne noise transfer, exhibiting a broadband character that dominates in the low- and mid-frequency range \citep{oboy2009tiremodel, koners2009airbornetire}. Apart from the broadband noise characteristic, the vibration due to the oscillation of air in the tire, referred to as tire cavity mode, is a tonal noise generation mechanism  \citep{bederna2009contributions}. In a recent study on electrified vehicle interior noise, this cavity mode was found to be the only perceptible tonal component of tire noise in the vehicle interior \citep{froehlingsdorf2023diss}.
The second noise source that exhibits broadband noise character is wind noise. Wind noise depends heavily on the driving speed and becomes a prominent contributor to the overall noise at high vehicle speeds \citep{cerrato2009windnoise,oettle2017automotiveaeroacoustics}. In \citep{Wickern2009Aeroacoustics}, the main sources of wind noise are considered to be leakage noise, cavity noise and the noise generated by separated flows at the outer surface. Concerning the aeroacoustic sources on the vehicle outer surface, authors distinguish between monopole sources due to unsteady volumetric flow, dipole sources characterized by fluctuating surface pressures at solid boundaries and quadrupole sources caused by unsteady shear stresses \citep{blumrich2006computational, oettle2017automotiveaeroacoustics}.

Apart from the aforementioned broadband sources, order-harmonic tonal noise from the electric drive unit and auxiliary components is audible in the vehicle interior as well as high-frequency inverter noise.
In \citep{Vijayraghavan1999emNoiseReview}, the authors review noise in electric machines and distinguish between magnetic, mechanical (load induced, auxiliaries, vibration modes), aerodynamic (air/ liquid /water cooling phenomena) and electronic (switching harmonics) noise sources. With a focus on the interior noise share of electric drive systems, high-frequency whining noise generated by electromagnetic forces and gear meshing of reducers as well as high-frequency inverter noise are particularly relevant \citep{kang2019electromagneticNoiseEM, mehrgou2018EMgearboxinv, froehlingsdorf2023diss}.

Beyond the electric drive unit, auxiliary components often contribute tonal, order-harmonic noise, particularly audible at low vehicle speeds \citep{patra2026HVAC, titave2025designHVAC}. Such auxiliary components include, but are not limited to, the electric refrigerant compressor, or  media-supplying components of fuel cell systems, namely the air compressor and the hydrogen recirculation blower. 

\subsection{Interior Noise Contribution Analysis}
Transfer path analysis (TPA) is an established family of test-based methods for identifying potentials for acoustic improvements in complex mechanical systems and is frequently applied for vehicle interior noise engineering. Distinguishing between source strength and corresponding transfer function, it allows engineers to target both -- sources and critical transmission paths. However, TPA methods like classical TPA and component-based TPA entail considerable effort: They require all relevant transfer paths from the noise sources to the receiver position to be instrumented with indicator sensors, typically microphones and tri-axial accelerometers \citep{tpa2016framework}. Operational measurements are then performed and complemented by additional explicit transfer path measurements, where the vehicle is in standstill condition. Operational Transfer Path Analysis (OTPA), first introduced in \citep{noumura2006otpa}, is frequently adopted as a lower-effort alternative in which the explicit quantification of source strengths and transfer functions is not the primary objective. Instead, OTPA delivers the contribution of each transmission path to the total interior noise from operational measurements only. While TPA can be described as a forward method that characterizes source strengths and transfer paths functions first, and then synthesizes the resulting interior noise contributions of the sources, OTPA can be seen as an inverse method. The measured receiver signal in the vehicle interior is an input to the algorithm, effectively being the quantity to be reconstructed by a linear time-invariant model, formed by multiplying the component-mounted indicator sensor signals with a transmissibility matrix that is fitted in the frequency domain in a least-squares sense. However, this approach has typical problems as analyzed in \citep{gajdatsy2008critical}, such as errors in estimation of transmissibility, errors occurring due to missing paths and errors occurring when indicator signals are strongly correlated.

An image-based line of work treats the separation of noise sources as a detection problem in which magnitude spectrograms are interpreted as images and the individual contributions are separated by detecting their characteristic patterns, in particular the frequency lines of tonal sources. Specifically for the interior noise of electric vehicles, the authors of \citep{froehlingsdorf2023atz, froehlingsdorf2023imeche, fvv1369} developed such methods for inverter noise, the tire cavity noise and electric drive unit noise, relying on the detection of straight lines in spectrograms and order-spectrograms via Hough transform. However, for the separation of the broadband sources wind and tire noise, an alternative mechanism to image-processing was applied, which is used as a reference baseline in this paper and is described in more detail in Section~\ref{sec:datagen}.
The detection of curved lines in spectrograms has been addressed for underwater acoustic signals with deep neural networks by the frequency line detection network (FLDNet) published in \citep{jiang2024fldnet}.

In \citep{prakash2022bayesiannvhmetamodelsassess, prakash2025stochastic}, the authors construct sophisticated probabilistic metamodels for electric vehicle interior noise. However, the goal is prediction in early product development phases, rather than contribution analysis. 

\subsection{Data-based Audio Source Separation}

Data-based audio source separation refers to the estimation of constituent signals from an observed mixture by models whose parameters are learned from data. The majority of these models are trained in a supervised fashion: the input to the models is the mixture and the output is the separated sources. During training, the reconstructed sources are compared to the existing ground truth source sounds for the respective mixture, which allows loss calculation and adjustment of the model parameters to improve the reconstruction. The field originated in speech and music source separation, and is related to data-driven audio enhancement, where speech or other target audio is to be removed from additive noise.
Recently, the field of source separation is being extended to universal separation models whose target is specified during inference.

Most early models were designed to operate in the spectrogram domain. Those models either pursue the prediction of a time-frequency mask that can be applied as a weighting function to the mixture spectrogram for source magnitude calculation \citep{Narayanan2013maskestimation, huang2014singing, jansson2017singing}, or the models pursue direct estimation of source spectrograms \citep{xu2013speechenancement, Stoeter2019, UhlichMusicDNN}. The resulting source spectrogram estimates can then be used in a multichannel Wiener filter \citep{Nugraha2016WienerPostDeep, Stoeter2019}, which combines them with the complex mixture spectrogram to obtain complex-valued source estimates while taking the spatial relationships between channels into account and promoting mixture consistency. Thus, the separated sources are jointly estimated such that their combination is encouraged to be consistent with the observed mixture. The separated source spectrograms are subsequently transformed back into the time domain using the inverse short-time Fourier transform (iSTFT).

In \citep{Luo2018TaSnet}, the authors discuss two drawbacks of Short-time Fourier Transform (STFT) to create a time-frequency representation of the sound: On one side, it is unclear whether the spectrogram domain provides the ideal signal representation for source separation. On the other side, the majority of proposed models discard estimation of phase information and use only the magnitude information, while reusing the phase of the mixture for the source estimates. This assumption is erroneous for each time-frequency bin where multiple sources contribute. The same authors published the popular time-domain -- also referred to as waveform-domain -- Conv-TasNet model \citep{luo2019convtasnet}. Further time-domain models are the Wave-U-Net \citep{stoller2018waveunet} and the waveform-domain Demucs \citep{defossez2019demucs} model for music source separation.

The Demucs successor model, called Hybrid Demucs, introduces a hybrid time-spectrogram architecture that can be described as two parallel U-Nets for parallel time and spectrogram branches, where the branches interact in their bottlenecks and after the inverse STFT of the spectrogram branch \citep{defossez2021hybrid}. A similar two-branched approach is taken in KUIELab-MDX-Net \citep{kim2021kuielabmdxnettwostreamneuralnetwork}. These hybrid time-spectrogram approaches reached state-of-the-art performance: to harmonize performance metrics of different separation models, a music demixing challenge \citep{MDC2021} was initiated. As model performance is strongly dependent on the dataset used for training, "Leaderboard A" was introduced to compare models that were all trained with the same MUSDB18-HQ dataset \citep{MUSDB18HQ}. The Hybrid Demucs model \citep{defossez2021hybrid} won this challenge, followed by KUIELab-MDX-Net \citep{kim2021kuielabmdxnettwostreamneuralnetwork}. In \citep{defossez2021hybrid}, the waveform and its STFT representation enter five parallel encoder layers that generate the same dimensions for both branches. In the following bottleneck, the representations are summed and passed to shared layers, before the branches are expanded back to their input dimensions. Skip connections from the respective encoder layers ensure the interpretability of the decoder branches still as spectrogram and time branch. The output of the spectrogram branch is processed by inverse STFT and summed with the waveform branch output, which gives the final model output. The Hybrid Transformer Demucs, HT Demucs \citep{rouard2023htdemucs} adapts its predecessor \citep{defossez2021hybrid} by replacing the innermost layers with a cross-domain transformer that applies self-attention within each domain and cross-attention across domains.

Another family of state-of-the-art models -- band-split recurrent and transformer models in the time-frequency domain \citep{luo2023bsrnn,lu2024bsroformer, wang2023melbandroformermusicsource} -- considers that a source's energy is typically concentrated in a limited frequency range. Rather than treating every frequency bin identically, which forces models to learn the respective frequency ranges for each source during training, band-split architectures use sub-bands seeking a more efficient representation. This proves effective: the Band-Split RNN (BSRNN) \citep{luo2023bsrnn} demonstrated good performance, and its transformer successors -- the Band-Split RoFormer \citep{lu2024bsroformer} and mel-scaled RoFormer \citep{wang2023melbandroformermusicsource} -- achieved state-of-the-art results on the standard benchmark \citep{MUSDB18HQ}. It is worth mentioning that these models are not exclusively using time-domain L1 loss (mean absolute error, MAE) computed between the reconstructed waveform and the ground truth waveform as done in hybrid models \citep{kim2021kuielabmdxnettwostreamneuralnetwork, defossez2021hybrid, rouard2023htdemucs}, but are including also a loss term measuring the complex spectrogram L1 loss. In \citep{lu2024bsroformer, wang2023melbandroformermusicsource}, the spectrogram loss is an additive multi-resolution spectrogram loss computed between the STFT of the prediction and the STFT of the target, evaluated at several different window sizes. Besides recent hybrid and band-split models, \citep{MDC2023} shows that using ensembles of different separation models has become a popular design choice.

A parallel line of research targets open-domain separation of arbitrary sounds. Early work addressed this in a query-free setting, separating mixtures of arbitrary sounds into class-agnostic outputs \citep{kavalerov2019universal}. Subsequent approaches condition the separator on a query that specifies the target: AudioSep \citep{liu2023audiosep} conditions separation on natural-language descriptions encoded by a contrastively pretrained text--audio model \citep{wu2023clap}, while the recent SAM Audio \citep{samaudio2025} accepts text, visual, and temporal-span prompts. It is a generative model built on a diffusion transformer architecture. Generative approaches to sound isolation are also popular in speech separation, audio enhancement and other text-prompted separation models \citep{Subakan2018GANsss, Zhu2022GenerativeFlow, mariani2024multisourcediffusionmodelssimultaneous, Scheibler2025flowmatching, Richter2023diffusionSpeechEnh, Yuan2025flowsep, Wang2025ExtractDiffTrans}.

\subsection{Conditioning mechanisms for separation}
The process of providing additional input, constraints or prompted context to a machine learning model is referred to as conditioning. A popular conditioning method for deep neural networks is feature-wise linear modulation (FiLM) \citep{perez2018film}, which conditions a network by applying a feature-wise affine transform whose scale and shift are predicted from side information. This has been used to steer separation networks toward a selected target, as in the Conditioned-U-Net \citep{meseguer2019cunet}. This model uses the spectrogram of a mixture as input and a control vector that indicates which instrument shall be isolated by the model. In \citep{choi2021conditioning-instruments}, the authors propose LaSAFT, which captures source-dependent frequency patterns and GPoCM, which seeks to modulate features more flexibly and expressively than FiLM. 

A distinct and, for the present work relevant form of conditioning supplies not the identity of the target source, but auxiliary side-information that is synchronized with the mixture and constrains the separation. This idea is established in music separation as score-informed separation, in which a time-aligned musical score parametrizes the decomposition \citep{miron2017monaural, gover2020score, tunturi2025scoreinformedmusicsourceseparation}. Likewise, singing voice separation based on the lyrics is implemented in \citep{jeon2020exploringalignedlyricsinformedsinging} and shows improvements of separation performance. However, as stated in \citep{tunturi2025scoreinformedmusicsourceseparation}, there is not much research on this form of conditioning deep learning models for music separation. Closest to the present application is the conditioning applied in \citep{gulli2025enhancing}, where speech enhancement for drone recordings is informed with the speed of the drone's motors. For three different speech enhancement models, the encoded rotor-speed features are either projected to the dimensions of the audio bottleneck and added, concatenated with the audio features at the bottleneck and processed jointly by a recurrent unit, or injected at multiple depths of the encoder.

\subsection{Design Choices and Contributions}

General-purpose and prompted separators offer greater flexibility but are trained on speech, music, and everyday sound events and are not exposed to large amount of training data from technical sound sources. Furthermore, the authors in \citep{samaudio2025} report that many prompted separation models improve flexibility, but frequently struggle within specialized domains. A supervised, domain-specific formulation is therefore preferred, as it is expected to achieve the desired performance within the target domain while remaining comparatively lightweight, owing to its narrower application scope and the reduced need to generalize across domains.

For the implementations in this work, the HT Demucs \citep{rouard2023htdemucs} model was selected. This is based on the following considerations:
(i) Music source separation can be seen as an application case closer to the technical domain than speech separation. Music instruments are harmonic sound sources covering a wide frequency range, whereas speech signals are usually more restrained in the frequency range. Furthermore, music source separation is often done for a set of 4 distinct sources, whereas many implementations for speech separation or speech enhancement are focusing on a smaller number of sources.
(ii) The authors published an officially maintained repository \citep{defossez_demucs_repo}.
(iii) In \citep{samaudio2025}, the authors benchmark their model against a selection of state-of-the-art models for music separation. HT Demucs \citep{rouard2023htdemucs} is considered by the authors as one of two open source models, while other benchmark models are proprietary and only accessible via public APIs. In this comparison, HT Demucs is scoring highest between the open-source models and close to the proprietary models in the overall subjective score introduced by the authors.

This paper makes the following contributions:
\begin{itemize}
  \item To the best of our knowledge, the first formulation of EV interior-noise decomposition as a supervised sound source separation problem
  \item A physics-based parametric generator that synthesizes labeled binaural source contributions from randomized virtual vehicle hardware and driving scenarios, applying domain randomization to promote transfer to real vehicles
  \item Adaptation of a state-of-the-art model for domain-specific sound source separation \citep{rouard2023htdemucs} for conditioning with operation conditions of the driving vehicle
  \item Separation results establishing a first estimate of achievable quality as a decision basis for future real-world data collection
  \item Suggestion of problem-related evaluation practice
  \item Discussion of conditioning for context-guided isolation of noise sources in technical systems
  \item An evaluation on measured interior noise of production electric vehicles, characterizing which source contributions of real vehicles can be inferred based on synthetic training data
\end{itemize}

The remainder of this paper is organized as follows. Section~\ref{sec:methodology} introduces the data generator, the conditioned architecture, the training objective, and the evaluation metrics, followed by the experimental setup and model variants in Section~\ref{sec:setup}. Section~\ref{sec:results} then traces the achievable separation quality. Sections~\ref{sec:discussion} and~\ref{sec:conclusion} return to the two questions posed in Section~\ref{sec:introduction} and derive the implications for future real-world data acquisition.
\section{Methodology}
\label{sec:methodology}

\subsection{Dataset}
\label{sec:datagen}

Due to the high cost of measurement campaigns associated with TPA and OTPA and due to the fact that the resulting data is usually treated as confidential, most organizations only have access to small, proprietary datasets. Therefore, the community has high interest in data-efficient training strategies, including synthetic and simulation-based data generation. Moreover, validation on inexpensive synthetic data allows the feasibility of the proposed approach to be assessed before substantial effort is undertaken with an unclear return on investment.

The data generation pipeline of this work proceeds in three stages. (i)~\emph{Virtual vehicle hardware:} a configuration is sampled, comprising wheel diameter, reducer gear ratio, pole-pair count, stator slot count, gear-tooth count, inverter switching frequency, tire-cavity-mode center frequency, and randomized binaural transfer functions from each of the four sources to the driver's left and right ear. Table~\ref{tab:random_property_distributions} provides an overview of the randomized values and value ranges. (ii)~\emph{Driving scenarios:} sections of $10\,\mathrm{s}$ duration are generated (accelerations, constant speeds, and decelerations), storing electric drive unit speed and load, vehicle speed, and auxiliary drive speed. (iii)~\emph{Component noise synthesis:} Each source contribution is synthesized based on the virtual vehicle hardware and the driving scenarios. Table~\ref{tab:operation_parameter_influence} summarizes the relationships between vehicle operation parameters and the affected noise sources. The overall broadband interior noise, composed of joint wind and tire contributions, follows the characterization given in \citep{bugl2025daga}, where a speed-dependent increase of broadband noise in electric vehicles is described. In \citep{fvv1369}, a method was developed to split the overall broadband interior noise into wind and tire noise. For each of 12 test vehicles, the ratio between the overall broadband interior noise and the OTPA-derived tire noise contribution was calculated at each frequency line. The median values across vehicles at each frequency line yielded a smoothed transfer function that predicts the tire noise contribution from the overall broadband noise alone. Once the tire noise is determined, the wind noise is derived as the result of the subtraction of the tire noise contribution from the overall broadband noise.
Leveraging the broadband noise model from \citep{bugl2025daga} and the split of broadband noise into wind and tire contributions from \citep{fvv1369}, the contributions for the two broadband sources are synthesized in this work. To vary the broadband noise generation of different virtual vehicles, the transfer function for the broadband noise split is slightly varied in the frequency domain. Furthermore, the wind noise amplitude is assigned a gain factor, which is randomized within a narrow range. As \citep{fvv1369} does not consider for the speed-dependence of the wind noise, an attenuation is introduced to reduce the generated wind noise amplitude below $50\,\mathrm{km/h}$, which is described in Table~\ref{tab:operation_parameter_influence}.

The powertrain (including inverter) and auxiliary drive noise contributions are introduced as tonal sources with varying orders and intensities. The generation of those sources was iteratively refined to yield a realistic overall acoustic impression. For the powertrain, the focus was to replicate the prominence of the whine noise over the broadband noise in the different speed regimes. Similarly, the prominence of the auxiliary drive was carefully designed to be only apparent to a low degree at low driving speeds.

All noise sources present in the vehicle interior are generated as mono signals, so that applying randomized transfer functions to the left and right driver ear yields the binaural source contributions.

The wide randomization of physically meaningful hardware parameters across plausible ranges is intended to make a real vehicle appear to the trained model as one further sampled configuration, so that the model generalizes beyond any single simulated setup. This use of broad parameter randomization in a simulator to promote generalization to unseen real conditions is the principle of domain randomization, introduced by \citet{tobin2017domainrand} for simulation-to-reality transfer in robotics and computer vision.

\subsection{Problem formulation}

Let $\mathbf{x}(t) \in \mathbb{R}^{2}$ denote the binaural mixture observed at time index $t$. The observed signal is modeled as the sum of $S$ different source contributions,
\begin{equation}
  \mathbf{x}(t)
  = \sum_{s=1}^{S} \mathbf{y}_{s}(t),
\end{equation}
with each $\mathbf{y}_{s}(t) \in \mathbb{R}^{2}$. In this work, $\mathbf{y}_s$ are the wind noise contribution, tire noise contribution, powertrain noise (electric drive unit and inverter) contribution, and auxiliary drive noise contribution. The separation model maps the binaural mixture $\mathbf{x}(t)$ to binaural estimates of all $S=4$ sources simultaneously.

\subsection{Network architecture and conditioning}
\label{sec:arc+cond}

The separation models is a hybrid time-spectrogram transformer of the HT Demucs type \citep{rouard2023htdemucs, defossez_demucs_repo}: a dual U-Net processes the signal in parallel in the waveform and spectrogram domains, and a cross-domain transformer encoder at the bottleneck applies self-attention within each domain and cross-attention between them, before symmetric decoders expand waveform and spectrogram representations per source. After inverse STFT of the spectrogram representations, the time-series data streams are summed to provide a waveform output per source. 

In addition to the acoustic mixture, we configure the model to receive time-synchronized operating-state information describing the current vehicle condition. The conditioning variables comprise the electric-drive rotational speed, drive load, vehicle speed, and auxiliary-drive rotational speed. These quantities are closely related to the physical excitation mechanisms of the individual source contributions. For example, vehicle speed governs the level of the broadband tire and wind noise and determines, together with the vehicle hardware parameters, the rotational speed of the electric drive; the latter determines the instantaneous frequencies of powertrain orders. Drive load additionally influences their amplitudes, while the auxiliary-drive rotational speed determines the corresponding auxiliary order frequencies. The operating-state signals therefore provide a physically grounded prior on the expected temporal, spectral, and amplitude characteristics of the individual source contributions.

The operating-state information is incorporated using FiLM~\citep{perez2018film}. We define the operating-state vector as
\begin{equation}
    \mathbf{o}(t) = \left[\, f_{\mathrm{drive}}(t),\ L(t),\ v(t),\ f_{\mathrm{aux}}(t) \,\right]^{T}
    \label{eq:operating_state}
\end{equation}
where $f_{\mathrm{drive}}$ denotes the electric-drive rotational speed, $L$ the drive load, $v$ the vehicle speed, and $f_{\mathrm{aux}}$ the auxiliary-drive rotational speed.

Because the different components of HT Demucs operate at different temporal resolutions, the operating-state signals are aligned to the temporal resolution of the respective feature representation before FiLM modulation. Let $\mathbf{h}_{l}(\tau) \in \mathbb{R}^{C_{l}}$ denote the feature vector at temporal position $\tau$ of layer $l$, where $C_{l}$ is the number of feature channels at that layer, and let $\mathbf{o}_{l}(\tau)$ denote the correspondingly aligned operating-state vector. A conditioning network $\varphi_{l}$ maps the operating-state vector to a pair of feature-wise scaling and shifting parameters,
\begin{equation}
    \left[\, \boldsymbol{\gamma}_{l}(\tau),\ \boldsymbol{\beta}_{l}(\tau) \,\right] = \varphi_{l}\!\left( \mathbf{o}_{l}(\tau) \right)
    \label{eq:film_params}
\end{equation}
with $\boldsymbol{\gamma}_{l}(\tau), \boldsymbol{\beta}_{l}(\tau) \in \mathbb{R}^{C_{l}}$. The corresponding FiLM transformation is
\begin{equation}
    \widetilde{\mathbf{h}}_{l}(\tau) = \boldsymbol{\gamma}_{l}(\tau) \odot \mathbf{h}_{l}(\tau) + \boldsymbol{\beta}_{l}(\tau)
    \label{eq:film_transform}
\end{equation}
where $\odot$ denotes element-wise multiplication. FiLM therefore allows the operating state to modify the activation of individual feature channels without requiring the conditioning variables to be concatenated directly with the acoustic representation.

Conditioning is introduced at multiple levels of the hybrid architecture. FiLM modulation is applied to intermediate representations in both the waveform and spectrogram branches, as well as to the representations processed by the cross-domain transformer. The conditioning information can therefore influence feature extraction independently in the two acoustic representations and subsequently their interaction in the transformer bottleneck.

\subsection{Training objective}

The HT Demucs model is trained with a source-weighted L1 reconstruction loss in the time domain \citep{rouard2023htdemucs}.
Let $\hat{y}_{s,b,c,t}$ denote the predicted waveform for source $s$, batch element $b$, audio channel $c$, and time index $t$, and let $y_{s,b,c,t}$ be the corresponding target (ground truth) waveform.
For each source, the absolute error is averaged over the $C$ audio channels, the $T$ time samples, and the $B$ batch elements:
\begin{equation}
  \ell_s^{\text{time}}
  =
  \frac{1}{BCT}
  \sum_{b=1}^{B}
  \sum_{c=1}^{C}
  \sum_{t=1}^{T}
  \left|
  \hat{y}_{s,b,c,t}
  -
  y_{s,b,c,t}
  \right|.
\end{equation}

The source-wise terms are then combined using per-source weights $w_s$:
\begin{equation}
  L_{\text{time}}
  =
  \frac{\sum_{s=1}^{S} w_s \, \ell_s^{\text{time}}}
       {\sum_{s=1}^{S} w_s}.
  \label{eq:time_weighted}
\end{equation}
The weights allow the training focus to be steered towards certain higher-weighted sources \citep{rouard2023htdemucs}. As the vehicle interior noise application involves source contributions with strongly varying amplitudes, the model tends to focus on the loudest source component, typically tire noise, while weaker sources are learned less effectively. To mitigate this imbalance, a source-wise loss normalization is introduced. In this formulation, the per-source waveform loss is divided by the mean absolute amplitude of the corresponding target source, with a small constant $\epsilon_{\text{time}} > 0$ to prevent division by zero. This brings the loss terms for different sources to a uniform order of magnitude and allows them to be combined with per-source weights $w_s$ as in Eq. \eqref{eq:time_weighted}, with $\ell_s^{\text{time}}$ replaced by $\ell_s^{\text{time,norm}}$. The normalization can optionally be enabled for training and is applied in the time domain:
\begin{equation}
\ell_s^{\text{time,norm}}=
\frac{1}{B}\sum_{b=1}^{B}
\frac{
\frac{1}{CT}\sum_{c=1}^{C}\sum_{t=1}^{T}
\left|\hat{y}_{s,b,c,t}-y_{s,b,c,t}\right|
}{
\max\left(
\frac{1}{CT}\sum_{c=1}^{C}\sum_{t=1}^{T}\left|y_{s,b,c,t}\right|,
\epsilon_{\text{time}}
\right)
}.
\end{equation}
To improve spectral source content, we add an auxiliary spectrogram-domain loss, using the same window size and hop length as the model's internal spectrogram branch. Let $Z_{s,b,m}$ denote the complex-valued STFT of the target source $s$ in batch element $b$, and $\hat{Z}_{s,b,m}$ the STFT of the corresponding estimate. The index $m \in \{1,\dots,M\}$ runs over the flattened time-frequency representation with $M = CFN$ bins, where $F$ is the number of frequency bins, $N$ the number of frames and $C$ the number of audio channels.

The magnitude spectrograms are denoted $|Z_{s,b,m}|$ and $|\hat{Z}_{s,b,m}|$.
Spectrogram loss can include up to three terms: a logarithmic magnitude term, a magnitude term, and a spectral convergence term, which follows the use in \citep{8528831}. These are defined as
\begin{equation}
  \ell_s^{\log}
  =
  \frac{1}{BM}
  \sum_{b=1}^{B}
  \sum_{m=1}^{M}
  \left|
  \log\left(|\hat{Z}_{s,b,m}| + \epsilon\right)
  -
  \log\left(|Z_{s,b,m}| + \epsilon\right)
  \right|,
\end{equation}
\begin{equation}
  \ell_s^{\mathrm{mag}}
  =
  \frac{1}{BM}
  \sum_{b=1}^{B}
  \sum_{m=1}^{M}
  \left|
  |\hat{Z}_{s,b,m}|
  -
  |Z_{s,b,m}|
  \right|,
\end{equation}
and
\begin{equation}
    \ell_s^{\mathrm{SC}}
    =
    \frac{1}{B}\sum_{b=1}^{B}
    \frac{
    \sqrt{\sum_{m=1}^{M}\left(|\hat{Z}_{s,b,m}|-|Z_{s,b,m}|\right)^2}
    }{
    \max\left(
    \sqrt{\sum_{m=1}^{M}|Z_{s,b,m}|^2},
    \epsilon_{\text{stft}}
    \right)
}.
\end{equation}
When multiple spectrogram loss terms are enabled, they are averaged with equal weights to form $\ell_s^{\mathrm{stft}}$. The source-weighted spectrogram loss is then
\begin{equation}
  L_{\mathrm{STFT}}
  =
  \frac{\sum_{s=1}^{S} w_s \, \ell_s^{\mathrm{stft}}}
       {\sum_{s=1}^{S} w_s}, 
\end{equation}
where the per-source weights $w_s$ are set to be identical in the time- and spectrogram-domain terms.
The final training objective is therefore
\begin{equation}
  \mathcal{L}
  =
  L_{\text{time}}
  +
  \lambda_{\mathrm{stft}}\,L_{\mathrm{STFT}},
\end{equation}
where $\lambda_{\mathrm{stft}}$ is the spectrogram loss weighting factor, a hyperparameter set before training.

\subsection{Evaluation metrics}
Separation quality is quantified with the signal-to-distortion ratio (SDR), which we adopt in line with the validation practice established in music separation. Following the formulation in \citep{MDC2021}, also referred to as \emph{nSDR} by \citet{defossez2021hybrid}, the time-domain definition of the SDR is:
\begin{equation}
  \mathrm{SDR}(\mathbf{y}_{s}, \hat{\mathbf{y}}_{s})
  =
  10 \log_{10}
  \frac{
    \sum_{t} \| \mathbf{y}_{s}(t) \|^2 + \epsilon
  }{
    \sum_{t} \| \mathbf{y}_{s}(t) - \hat{\mathbf{y}}_{s}(t) \|^2 + \epsilon
  }.
\end{equation}
This formulation of SDR is not to be confused with a prior definition of SDR, published in \citep{vincent2006bsseval}. Since this time-domain metric is not fully representative of separation quality \citep{samaudio2025}, the SDR values reported in the following are complemented by a spectral inspection of the separated signals.

\section{Experimental Setup}
\label{sec:setup}

For the experiments, two datasets are generated. The first dataset consists of a large number of driving conditions sampled for a single virtual vehicle. This ensures that the hardware parameters, including the binaural transfer functions from sources to interior, the number and intensity of powertrain orders, and the prominence of the auxiliary drive, are kept constant. This allows for simplified experimentation on model variants and conditioning. In addition, a second dataset is generated, which includes $300$ different virtual vehicles with variations in the hardware, that the model needs to generalize across. This dataset is also used to investigate whether the noise contributions of real vehicles can be separated using the model trained on virtual data.

The experiments use the data generation pipeline described in Sec.~\ref{sec:datagen}. All source and mixture signals are binaural, $10\,\mathrm{s}$ long, and generated at $44.1\,\mathrm{kHz}$. During training, audio is resampled to $16\,\mathrm{kHz}$, and training samples are formed with a segment length of $6\,\mathrm{s}$ and a shift of $2\,\mathrm{s}$, yielding three samples per generated driving scenario. 
For the presented experiments, we filter for accelerated driving scenarios with a maximum speed of $100\,\mathrm{km/h}$. This corresponds to $9237$ / $93$ training/ validation samples for the data set with a single virtual vehicle and to $6093$ / $2127$ training / validation samples for the data set with multiple virtual vehicles.

All models use the same conditional HT Demucs backbone (Sec.~\ref{sec:arc+cond}) with three encoder-decoder stages in both the waveform and spectrogram branches. The initial feature channel width is $24$ in each branch, growth factor $2$, kernel size $8$, and stride $4$. Each encoder stage includes one DConv residual unit (DConv depth $1$, compression $4$, with decoder-side DConv disabled), and the bottleneck is a 2-layer cross-transformer with $4$ attention heads, hidden scale $4$, and sinusoidal positional embeddings. This configuration effectively provides a lightweight version of the original model \citep{rouard2023htdemucs}. Where conditioning is enabled, it is injected by a 2-layer FiLM network with hidden width $16$ at frame, time-branch, and transformer levels. The Demucs-internal data augmentations are disabled to preserve audio--conditioning alignment and normalization of audio, that enters the model is disabled to preserve the relation between vehicle operating state (e.g. vehicle speed) and noise levels.

The window size of the STFT is $4096$ and the hop length is set to $1024$ unless stated otherwise. The Demucs repo \citep{defossez_demucs_repo} offers two different spectrogram representation modes. The default is Complex-as-channels (CaC) mode, where complex STFT values are represented as channel pairs (real/imaginary) and predicted directly by the network. Inverse STFT is then applied without iterative Wiener refinement. Alternatively, the complex STFT can be estimated with single-channel representation, so the model predicts magnitude-domain source estimates that are converted to complex source STFTs via iterative Wiener filtering. In this study, we disable the complex-as-channels mode and use magnitude prediction followed by Wiener-based refinement using one refinement step in training and one refinement step in inference. The optional residual branch in this refinement stage is disabled, so mixture energy is assigned only to the modeled sources. This setup corresponds to a light Wiener refinement focused on mixture-consistent source reconstruction, which is not always a constraint enforced for sound separation models. 

Optimization uses Adam with initial learning rate $5\times10^{-4}$, $\beta_1=0.9$, $\beta_2=0.999$ and no weight decay. Training uses batch size $8$, up to $120$ epochs, random seed $42$, and $10$ data-loader workers.

For waveform reconstruction, per-source weights are set to $w_s=[1,1,20,20]$ for tire, wind, powertrain/inverter, and auxiliary drive in order to enhance the separation performance for the heavily masked contributions. Loss-normalized variants use $\epsilon_{\text{time}}=10^{-6}$. For variants with auxiliary spectrogram loss, the reported runs use spectral-convergence-only STFT loss (log-magnitude and magnitude terms disabled), with $\epsilon_{\text{STFT}}=10^{-8}$ and $\lambda_{\mathrm{STFT}}=1$.

All experiments are run on a single node of an HPC cluster. Each job is allocated one NVIDIA H100 GPU (96\,GB HBM2e memory), eight cores of an Intel Xeon 8468 Sapphire CPU, and $122\,\mathrm{GB}$ of system memory.

To isolate the contributions of conditioning, loss normalization, and spectrogram loss, eight model variants are trained and evaluated for the dataset composed of a single virtual vehicle. The variants are summarized in Table~\ref{tab:experiment_variants}. Apart from the explicitly ablated factors, other architecture and optimization hyperparameters are held fixed.

\begin{table}[htbp]
\centering
\caption{Overview of model variants differing in conditioning, waveform-loss normalization, and use of an auxiliary spectrogram loss.}
\label{tab:experiment_variants}
\small
\begin{tabular}{l c c c p{5.2cm}}
\toprule
\textbf{Abbrev.} & \textbf{Conditioning} & \textbf{Loss Norm.} & \textbf{$L_{\mathrm{STFT}}$} & \textbf{Description} \\
\midrule
UC-BASE     & --   & --  & --  & Baseline: unconditioned, unnormalized waveform loss, no spectrogram loss\\
UC-N     & --   & \checkmark & --  & Unconditioned with normalized waveform loss \\
UC-STFT   & --   & -- & \checkmark & Unconditioned, unnormalized waveform loss,  with auxiliary spectrogram loss \\
UC-N-STFT   & --   & \checkmark & \checkmark & Unconditioned, normalized waveform loss, with auxiliary spectrogram loss \\
FiLM     & FiLM & --  & --  & FiLM-conditioned, unnormalized waveform loss \\
FiLM-N   & FiLM & \checkmark & --  & FiLM-conditioned with normalized waveform loss \\
FiLM-STFT & FiLM & -- & \checkmark & FiLM-conditioned, unnormalized waveform loss, with auxiliary spectrogram loss \\
FiLM-N-STFT & FiLM & \checkmark & \checkmark & FiLM-conditioned, normalized waveform loss, with auxiliary spectrogram loss \\
\bottomrule
\end{tabular}
\end{table}

\section{Results}
\label{sec:results}

\begin{figure}[htbp]
    \centering
    \includegraphics[width=\textwidth]{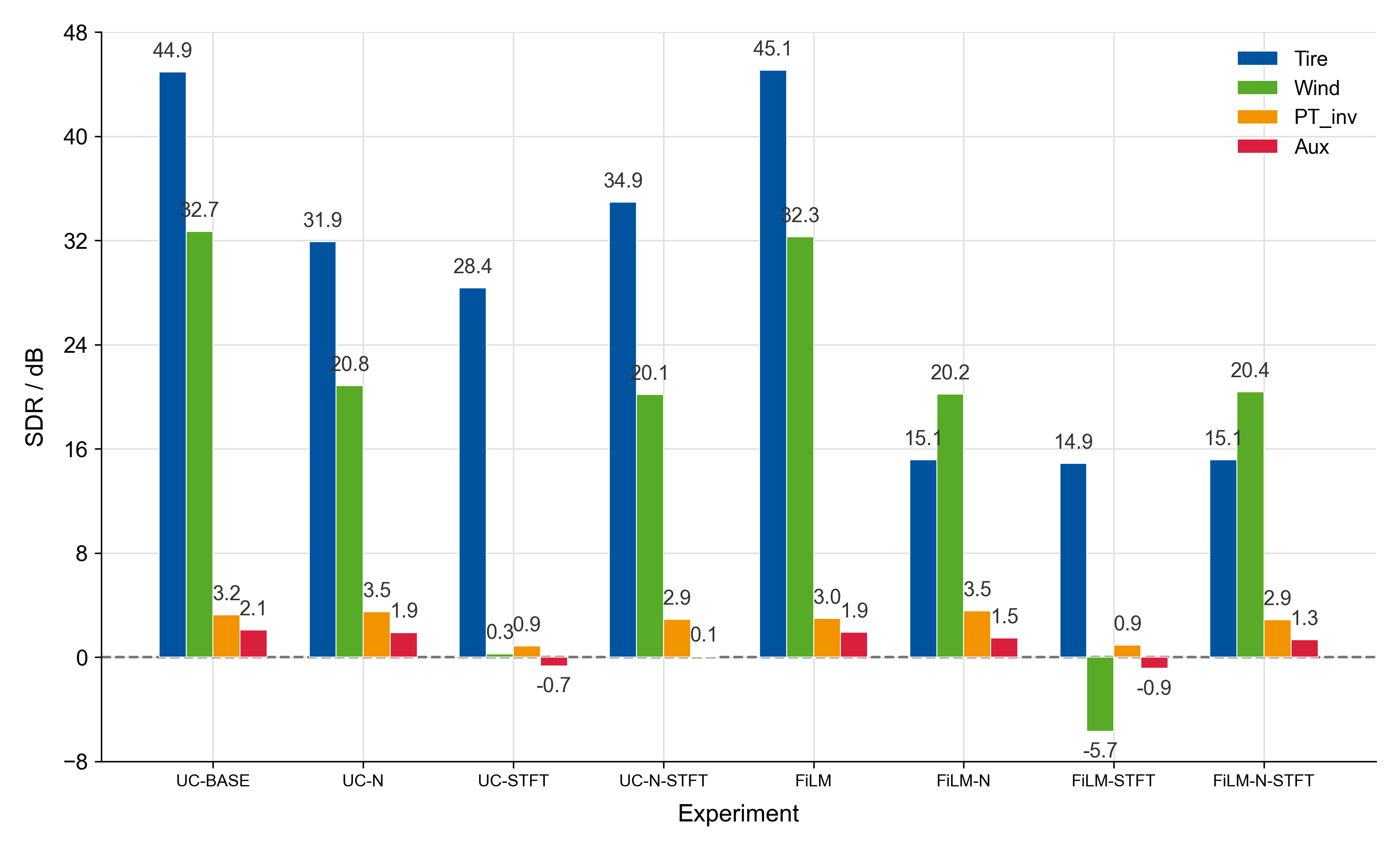}
    \caption{SDR separation results for single virtual vehicle.}
    \label{fig:singleVeh_accel_weights20_nfft4096_hop1024_wiener1_norm-off_detail}
\end{figure}

As illustrated in Fig.~\ref{fig:singleVeh_accel_weights20_nfft4096_hop1024_wiener1_norm-off_detail}, all model variants achieve high SDR values for the separation of wind and tire noise. These broadband sources are the main contributors to the overall vehicle interior noise and their high SDR values indicate that their time-series reconstruction functions effectively. Naturally, normalizing the losses for the different sources shifts the focus of the separation towards sources with lower amplitudes. Concerning the SDR, this is more apparent as a reduction of the tire and wind noise SDR, than as an increase in powertrain and auxiliary drive SDR. Powertrain and auxiliary drive show a comparatively low SDR, which is expected, as the source content is strongly masked. Variants using spectrogram loss in addition to time-domain loss score low in SDR, because the optimization goal of these models is not only the time-series reconstruction, but also spectrogram reconstruction. Therefore, in the time-series based SDR metric, these models are likely to score lower. Inspecting the conditioned variants, it was found that the SDR is not a suitable indicator for separation quality in this application. In several cases, estimates with favorable SDR values exhibited poor spectral quality, including artifacts such as periodically recurring bursts of energy along the time or frequency axis. Therefore, the evaluation should complement SDR with auditory inspection and spectrogram analysis.

\begin{figure}[htbp]
    \centering
    \includegraphics[width=\textwidth]{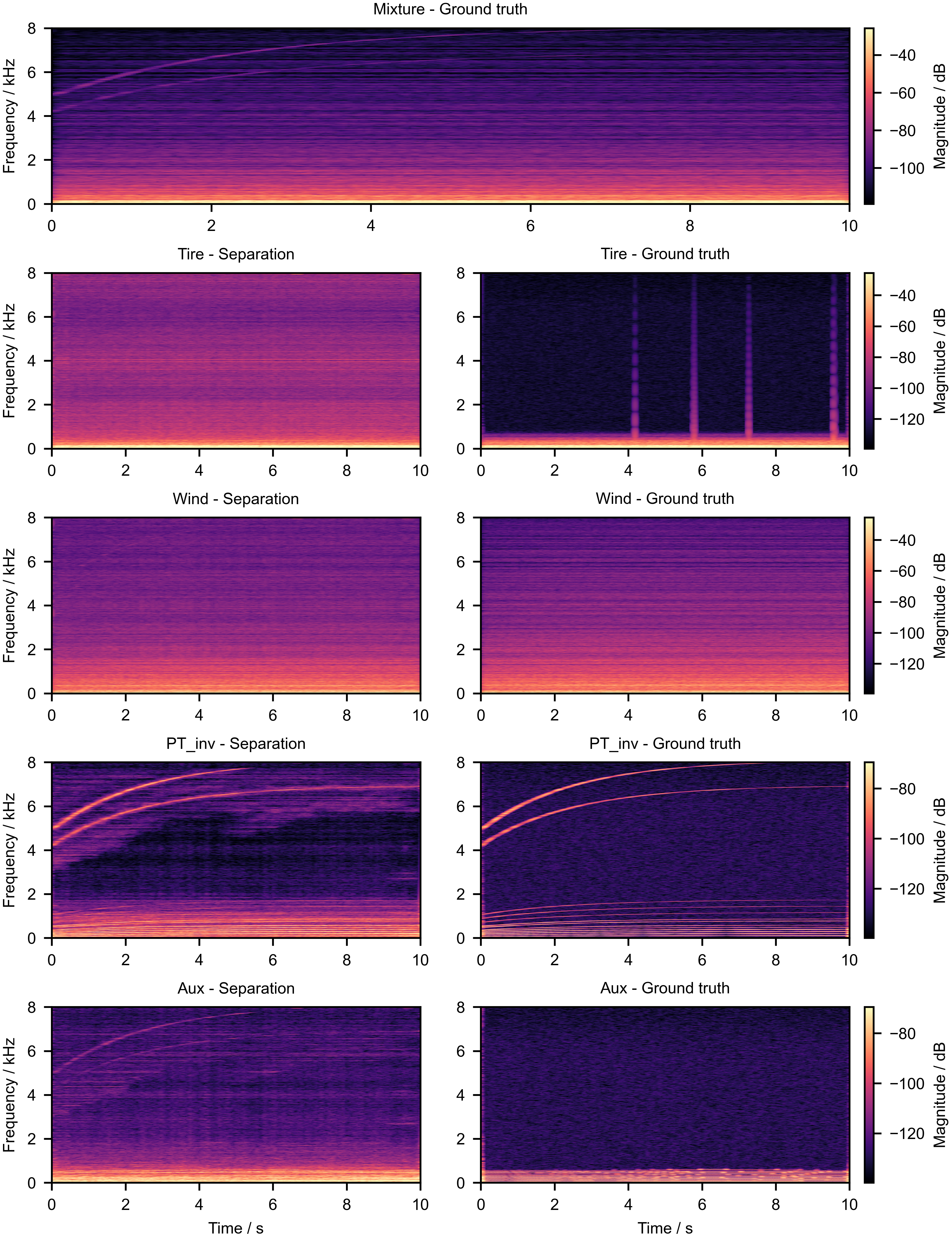}
    \caption{Spectrograms for separation of source contributions of a single virtual vehicle. Experiment UC-STFT.}
    \label{fig:seed_1_sample_0_1383_gtmixrow}
\end{figure}

Fig.~\ref{fig:seed_1_sample_0_1383_gtmixrow} shows the spectrogram of the mixture in the top row and compares the estimated source contributions with their corresponding ground-truth signals. Note that the noise levels of the individual sources were calibrated relative to one another during data generation, but not with respect to absolute sound pressure levels; the decibel values in the following shown should therefore be interpreted on a relative scale only. The example illustrated here was obtained using the best-performing model variant. The tire noise, characterized by its low-frequency content, is approximated very well, although visual inspection reveals an elevated noise floor above $1\,\mathrm{kHz}$. This is because the loss is mainly defined by frequency regions with large amplitudes. Consequently, accurate reconstruction of the high-energy low-frequency tire-noise content is emphasized, whereas errors at higher frequencies have a comparatively small influence. In contrast, the spectral energy of wind noise is distributed more broadly across frequencies, resulting in separation focus being broader in the frequency range.  Its spectral content is reproduced particularly well below $2\,\mathrm{kHz}$, which indicates that the wind noise contribution, shaped by the binaural transfer functions, is internalized by the model. The powertrain contribution is weak in this example: only two orders are visible above $4\,\mathrm{kHz}$ as curved lines. These curved order trajectories are estimated accurately at the beginning of the measurement, when the electric-drive load is higher during vehicle acceleration. As acceleration decreases, the orders approach approximately constant frequencies and become less prominent. Fully masked harmonic content, such as order lines below $2\,\mathrm{kHz}$ cannot be recognized with image-based separation methods. Nevertheless, the model predicts lower-order components in this frequency range, but separates them less clearly than the less-masked orders. For the auxiliary drive noise contribution, the model reproduces the expected low-frequency concentration and low-magnitude narrowband order content, but the estimate remains less accurate than for the other sources.

\begin{figure}[htbp]
    \centering
    \includegraphics[width=\textwidth]{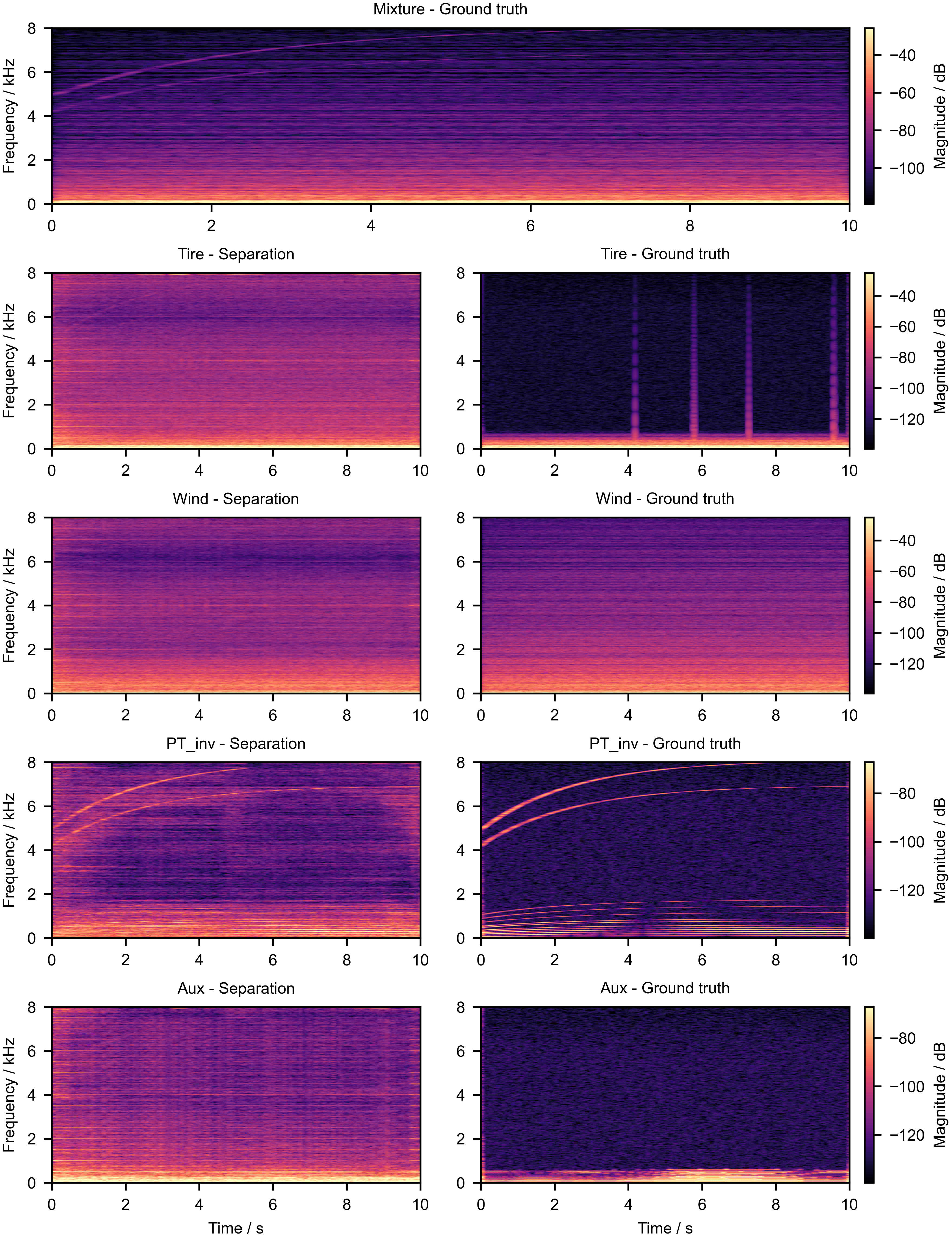}
    \caption{Spectrograms for conditioned separation of source contributions of a single virtual vehicle. Experiment FiLM-N-STFT.}
    \label{fig:seed_1_sample_0_1383_gtmixrow_cond}
\end{figure}

The conditioned variants produce lower separation quality than the unconditioned variants, despite favorable SDR values for some configurations. In principle, the additional conditioning channels provide information that should support source separation. However, their effective integration into the network proved challenging. The best conditioned-model performance was obtained with an STFT window size of $8192$ and hop size of $256$ samples in the spectrogram branch. Because a smaller hop size substantially increases the number of spectrogram frames and, consequently, the computational cost and training time, $256$ samples represented the smallest practically feasible value for the available computing setup. Fig.~\ref{fig:seed_1_sample_0_1383_gtmixrow_cond} shows results that are inferior to those in Fig.~\ref{fig:seed_1_sample_0_1383_gtmixrow}, particularly with respect to the isolation of the powertrain contribution. Moreover, the conditioned estimates exhibit horizontal bands of increased energy at intervals of $250\,\mathrm{Hz}$. This interval corresponds to the sampling rate of the time-branch feature channels that are provided to the cross-domain transformer encoder. Furthermore, the effective frequency resolution of the spectrogram branch at the bottleneck is $250\,\mathrm{Hz}$ for the STFT window size of $4096$ ($125\,\mathrm{Hz}$ for window size of $8192$). Conditioning was investigated through injection exclusively into the transformer, the spectrogram encoder layers, and the time-domain encoder layers. However, none of these configurations eliminated the observed artifacts. Reducing the STFT hop size was the most effective mitigation measure. A smaller hop size increases the temporal resolution of the spectrogram representation, while also increasing computational cost, and appears particularly beneficial for conditioned models because it enables a more temporally consistent alignment between the spectrogram features and the conditioning information.

\begin{figure}[htbp]
    \centering
    \includegraphics[width=\textwidth]{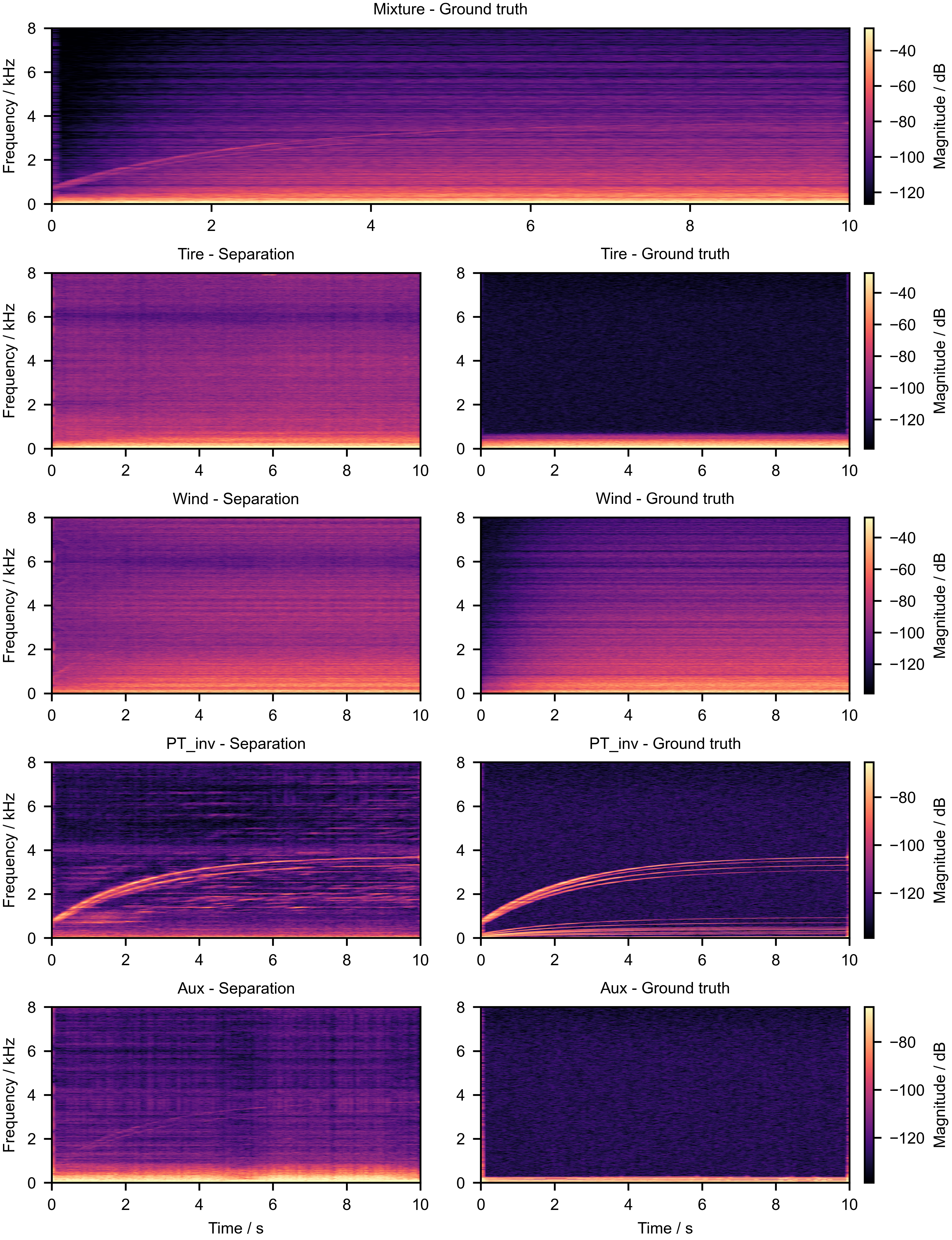}
    \caption{Spectrograms for separation of source contributions for one of multiple virtual vehicles. Experiment UC-STFT.}
    \label{fig:seed_10_sample_0_11_gtmixrow}
\end{figure}

As the final goal is to perform sound source separation robustly even for vehicles not seen during training, our experiments also evaluate a model trained on multiple virtual vehicles. This setting poses the considerably greater challenge: the sources are not only spectrally overlapping, but also differ in character across vehicles, exhibiting distinct transfer functions to the interior, different order spectra, and varying relative intensities. Fig. ~\ref{fig:seed_10_sample_0_11_gtmixrow} shows the separation results obtained with the best-performing model in this multi-vehicle setting. While tire and wind noise are approximated with very good accuracy, the powertrain noise reveals slight artifacts and the model is not capable anymore to infer the masked low-frequency order content. The auxiliary drive noise is likewise harder to infer, as it is not only completely masked by the other sources, but also showing larger variations in the training data. However, it should be noted that predicting very low noise levels -- even with hallucinated spectral content -- may be considered a partial success.

\begin{figure}[t]
    \centering
    \begin{subfigure}[t]{0.49\textwidth}
        \centering
        \includegraphics[width=\textwidth]{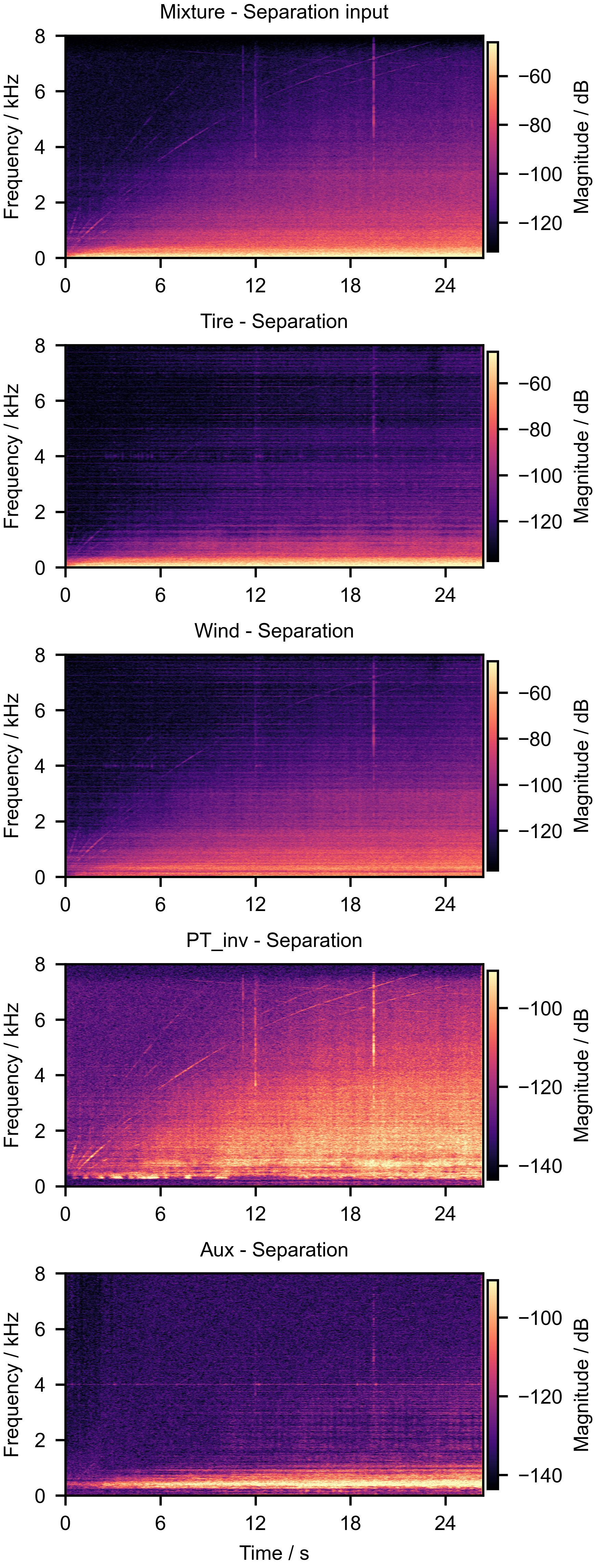}
        \caption{Vehicle A - Crossover SUV}
        \label{fig:sub-a}
    \end{subfigure}
    \hfill
    \begin{subfigure}[t]{0.49\textwidth}
        \centering
        \includegraphics[width=\textwidth]{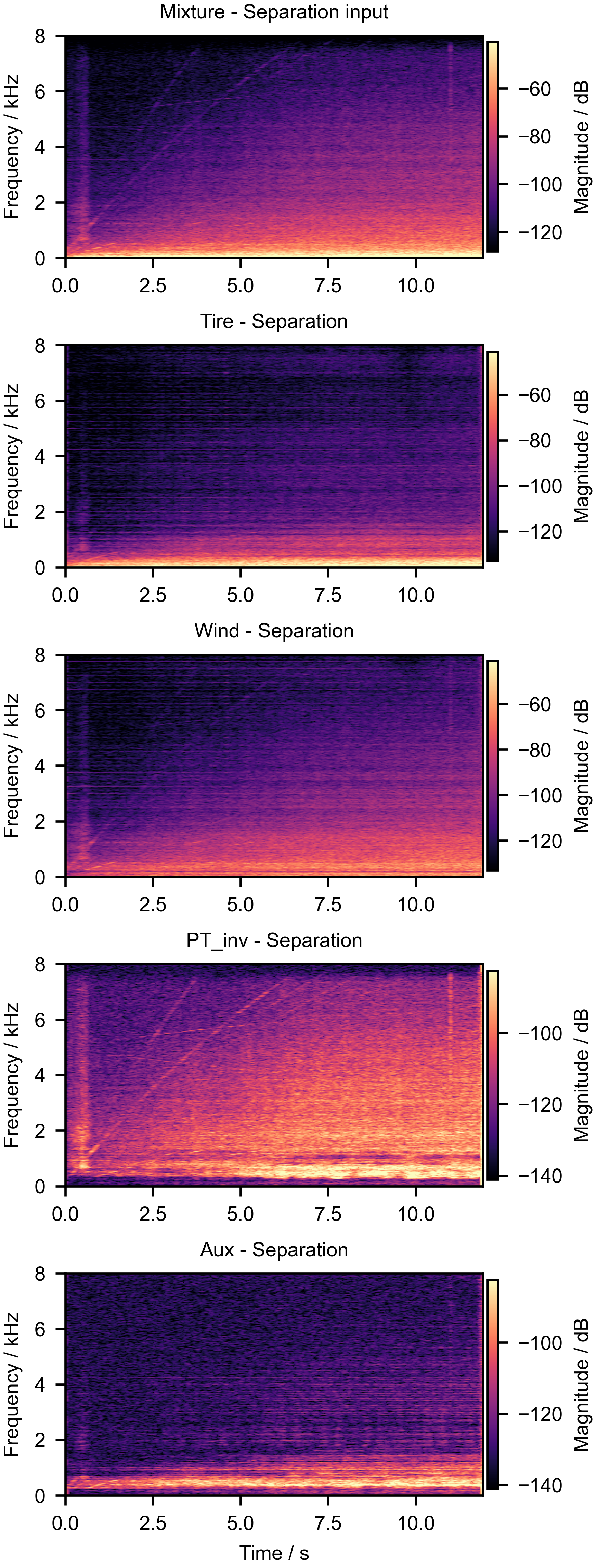}
        \caption{Vehicle B - Hatchback}
        \label{fig:sub-b}
    \end{subfigure}
    \caption{Source contribution spectrograms for real vehicle interior noise measurements recorded in \citep{fvv1369}. As the mixtures are normalized before inference, the magnitude scale is relative and does not reflect measured decibel values.}
    \label{fig:comparison}
\end{figure}

Fig.~\ref{fig:comparison} shows the performance of the separation model on real vehicle-interior noise measurements. The best-performing variant is UC-BASE, as it is able -- in contrast to the UC-STFT variant -- to assign inverter switching noise to the powertrain and to more efficaciously allocate powertrain orders. However, results remain unsatisfactory overall. Particularly during the initial phase of the accelerated measurements, order-harmonic noise erroneously appears in the wind- and tire-noise estimates. Nevertheless, most of the energy in the audible orders is correctly attributed to the powertrain. The inverter noise is likewise correctly attributed to the powertrain -- its switching frequency is recognizable at approximately $7.5\,\mathrm{kHz}$ for Vehicle A and at $5\,\mathrm{kHz}$ for Vehicle B. At higher vehicle speeds, the model attributes noise to the powertrain across a broad frequency range. This results from lower-order content being present in the training data -- content that is also present in real vehicles, but it is masked by tire and wind noise and therefore imperceptible -- causing the model to hallucinate this content. Likewise, the source content of the auxiliary drive noise is hallucinated.

\section{Discussion}
\label{sec:discussion}

\subsection{Feasibility and transfer}
\label{sec:disc_feasibility}
The separation task posed here is a demanding one. All four contributions are active at every instant, they are driven by a common operating state, and the tonal contributions of powertrain and auxiliary drive are masked by the broadband contributions of tire and wind noise over most of the frequency range. Additionally, the evaluated scenarios exclude full-load accelerations, during which the powertrain is most strongly excited and its order content is least masked.

Under these conditions, the experiments on a single virtual vehicle show that the decomposition is learnable. Tire and wind noise are reconstructed with high fidelity, and tonal powertrain content below $2\,\mathrm{kHz}$ is recovered although it is heavily masked in the mixture. Training with the dataset containing different virtual vehicles with variations in the hardware degrades the results in a gradual and source-dependent manner rather than abruptly: tire and wind noise remain satisfactory, the powertrain estimate exhibits artifacts and loses its low-frequency order content, and the auxiliary drive, which is both completely masked and subject to large variation across vehicles, is affected most. The first question posed in Section~\ref{sec:introduction} can therefore be answered affirmatively for the synthetic domain, with the qualification that separation quality is governed by the excitation level of a contribution relative to the broadband masking.

The transfer to measured vehicle interior noise is partial and selective. Order-harmonic content appearing erroneously in the wind and tire estimates and hallucinated powertrain content at higher vehicle speeds are consequences of the training distribution. The simulation-to-reality gap is thus not closed by domain randomization of the virtual vehicle hardware alone. Since the observed failures are systematic rather than arbitrary, and since they concern the distribution of the training data rather than the capacity of the separator, the more promising route is not a further refinement of the generator, but the use of measured training data with known source contributions.

\subsection{Implications for the training objective, evaluation, and model configuration}
\label{sec:objective_eval_model-config}
The training objectives established in music demixing are coupled to its evaluation practice: the SDR metric of the music demixing challenges \citep{MDC2021, MDC2023} is a time-domain energy ratio, so a purely waveform-based loss is the most direct means of scoring well. In the present application, the waveform reconstruction is valuable, which is proven through well-functioning variants trained exclusively with the time-domain loss. However, as the quantity of interest are good contribution results, where the spectral compositions of the contributions are free from misallocated signal content, the introduced auxiliary spectrogram loss is justified and beneficial.

As the per-source loss is dominated by signal content with large values, for the tire contribution this emphasizes the high-energy low-frequency content, which is reconstructed accurately, while the elevated noise floor above $1\,\mathrm{kHz}$ contributes too little to the loss to be corrected. The remedy of a logarithmic magnitude term, however, was found to overcorrect: training with the log-magnitude term shifted the emphasis towards low-level background content and underfitted the high-energy regions, resulting in poor overall separation. The multi-resolution spectrogram loss employed in \citep{lu2024bsroformer, wang2023melbandroformermusicsource} seems to be a natural extension of spectrogram-based objectives, as it incorporates multiple time-frequency resolutions during loss computation.

A similar reasoning applies to the evaluation of the results. Just as waveform reconstruction should not be the only optimization goal, the time-domain SDR is not to be considered a sufficient metric for performance evaluation of technical noise sources. Its limitations are evident in the results in two ways: variants trained with an auxiliary spectrogram loss score lower in SDR while producing spectrally cleaner source estimates, and several conditioned variants attain favorable SDR values while exhibiting periodically recurring bursts of energy along the time or frequency axis that are immediately apparent in the spectrogram and clearly audible. The error of the third-octave band levels for the broadband contributions and the level error along tracked order trajectories for the tonal contributions are therefore suggested as complementary metrics, which both are compatible with interpretations established in acoustic engineering. Both of those can also be considered in the loss function. However, this requires detailed consideration of their properties to ensure that backpropagation remains possible. 

Two configuration choices proved to be beneficial independently of the exact training configuration. The phase of an individual contribution in this setup is of minor importance, since the sources are largely mutually uncorrelated. Predicting magnitudes and recovering the complex estimates by Wiener filtering is therefore an acceptable trade, and it additionally promotes mixture consistency: with the residual branch disabled, the energy of the mixture is distributed among the modeled sources, so that any energy assigned to one contribution is withdrawn from another. This procedure proved useful for suppression of hallucinated source content. Furthermore, since the tonal contributions have very narrowband character, a fine frequency resolution in the spectrogram branch is beneficial for the powertrain estimate. This motivated the choice of the reported window size of the STFT.

The conditioning with the operating state, in contrast, did not meet expectations. Although the operating state is physically informative about where and with what intensity source-specific content is to be expected, the conditioned variants did not improve upon the unconditioned ones and were spectrally inferior. The characteristic artifact is a set of horizontal bands spaced at $250\,\mathrm{Hz}$, which coincides both with the sampling rate of the time-branch features at the bottleneck and with the frequency resolution of the spectrogram branch at the bottleneck. Restricting the injection to the transformer, to the spectrogram encoder, or to the time-domain encoder did not significantly reduce the bands, whereas reducing the STFT hop size did, because of the associated higher rate of condition injection. Two aspects remain unexplored. Only the injection point was varied, not the representation of the conditioning vector itself; the four variables differ in unit and numerical range and their relation to the resulting source amplitudes is nonlinear, which motivates an investigation of a suitable representation. In addition, the conditioning is possibly redundant in the single-vehicle setting, since vehicle and drive speed can presumably be inferred from the mixture itself -- the expected benefit is greatest where the model is currently weakest, namely for unseen vehicle hardware and for masked order content.

\subsection{Relation to established methods and applicability}
\label{sec:disc_relation}
Once trained, the model requires only a binaural measurement at the receiver position, complemented by the operating-state signals if the conditioning is to be used. Compared with TPA, which requires instrumentation of all relevant transfer paths and additional measurements at standstill, and with OTPA, which still requires instrumentation of the transfer paths, the measurement effort is reduced substantially, and vehicles that cannot be instrumented become accessible to a contribution analysis. The upsides of TPA and OTPA campaigns on the other side are a higher accuracy, and the opportunity of ranking transmission paths. TPA further allows quantification of source strengths and transfer functions, making it an indispensable tool for evaluating design changes. The proposed method cannot replace these capabilities, but its substantial time savings and ease of application justify its position as a third alternative alongside TPA and OTPA, particularly where instrumentation is impractical or infeasible. Beyond serving as an alternative, the method can also be combined with OTPA: contributions whose instrumentation is particularly costly, such as tire noise, could be estimated by the learned model, while OTPA covers the remaining paths, keeping OTPA campaigns feasible even when not all sources can be equipped with indicator sensors.

The proposed approach differs from the image-based separation methods developed for this application \citep{froehlingsdorf2023atz, froehlingsdorf2023imeche, fvv1369} in three aspects. It recovers order content that is heavily masked in the mixture and therefore not detectable as a line in the spectrogram, provided that no generalization across vehicle hardware is required. Furthermore, the decomposition of the broadband noise into tire and wind contributions is learned from the data distribution rather than imposed: in \citep{fvv1369}, the decomposition relies on a frequency-dependent ratio determined as a median across test vehicles and is thus identical for every vehicle and operating point, whereas a learned model can represent speed-dependent differences once it is trained on measured data. While authors in \citep{fvv1369} carefully crafted the methods to differ between the contributions of different components, the proposed method omits manual feature engineering and relies on data instead -- which offers scalability to incorporate and learn from new data points with low manual effort.

The considerations also point to further application cases for the method. The decisive criterion for the applicability of the method is the availability of training data in which the individual source contributions are known. Three situations can be distinguished. In the most favorable one, each source can be operated and recorded in isolation at the receiver position, and mixtures are formed by summation, so that labeled training data is obtained without ever measuring a mixture and without a transfer path analysis. In the second, the sources cannot be isolated in the complete system, but they can be reproduced separately on partial test rigs. Only in the third, which is the case addressed in this work, do the contributions have to be recovered from operational measurements by a transfer path analysis, which is the reason for the high cost of the required data and, ultimately, the motivation for the synthetic data generator. Independently of this, the inference side is always the same: once the model is trained, a single measurement at the receiver position is sufficient.

Within automotive engineering, the first situation is encountered whenever the vehicle is stationary and the broadband contributions of tire and wind noise are absent, so that the auxiliary components determine the interior sound. This applies to cabin conditioning through the thermal-management system and, most prominently, to DC fast charging, during which the refrigerant compressor, the coolant pumps and the fans operate at high load while the occupants are inside or next to the vehicle. These components can be activated individually through the vehicle control unit, their shaft speeds are available as conditioning information, and the resulting separation task involves only few, comparably loud contributions. The second situation is represented by wind noise measured in the wind tunnel, tire noise on a drum rig, and powertrain noise on a dynamometer, each recorded separately for the same vehicle and subsequently summed. Furthermore, vehicle pass-by noise, for which isolated tire-noise contributions are frequently measured on dedicated test vehicles with the remaining sources suppressed, can be an application case. The ease of use enables for further application cases, such as analysis of competitor vehicles in a low-effort benchmarking campaign. An example outside of vehicle engineering is the noise in production halls, in which machines can be operated individually during commissioning or maintenance, whereas the attribution of the noise exposure at a workplace is of interest during production.

\section{Conclusion}
\label{sec:conclusion}
This work formulates the decomposition of electric vehicle interior noise into the contributions of wind, tire, powertrain and auxiliary drive as a supervised source separation problem and addresses it with conditioned and unconditioned variants of a hybrid time-spectrogram transformer trained on physics-based synthetic data.

A first outcome concerns the assessment of such a decomposition. The time-domain SDR inherited from music demixing proved insufficient as a criterion in this domain, as favorable values were repeatedly obtained for estimates whose spectral content was evidently deficient. An inspection of the spectrograms and an auditory assessment of the separated contributions are therefore indispensable, and they should be complemented by metrics established in acoustic engineering, such as the error of the third-octave band levels for the broadband contributions.

The main finding is that a data-driven approach is effectively able to learn the sound source separation task for given technical noise sources. Whenever a source contributed sufficiently to the interior noise to be audible or visible in the spectrogram, the separation model effectively isolated this contribution. For measured vehicle interior noise, however, the synthetic data generated by domain randomization is not sufficient on its own. Instead, it can provide a pre-trained model for subsequent fine-tuning on measured data, which may reduce the amount of measurement data required.

Conditioning with the vehicle operating state did not yet yield the expected benefit and is consequently identified as the primary direction for further research, in particular with respect to the selection and the representation of the conditioning variables and to the manner of their injection into the separation model.

Once trained, the method requires no more than a measurement at the receiver position. It therefore complements transfer path analysis and its operational variant as a third, low-effort alternative, applicable wherever labeled source contributions can be obtained.

\section*{Acknowledgements}

This publication is a scientific result of the research project \textit{ML/AI Support for NVH Engineering, FVV project no. 1555} funded by the FVV eV and jointly performed by the Chair of Thermodynamics of Mobile Energy Conversion Systems at RWTH Aachen University, the Chair of Vibroacoustics of Vehicles and Machines at the Technical University of Munich, and the Data Analytics and Machine Learning group at Forschungszentrum Jülich. The project was accompanied by a group of experts led by Dr. Michael Fischer (Robert Bosch GmbH). We gratefully acknowledge the support received from the project coordination and from all members of the project committee.

Computational resources were provided by the German AI Service Center WestAI.
    
During the preparation of this work, the authors used GitHub Copilot in order to assist with coding implementation, including the conditioning mechanism added to the HT Demucs model, code modifications, and repository documentation. The authors used Perplexity AI in order to (i) support the development of the physics-based parametric data generation script, and (ii) assist in identifying related work during the literature review. The authors also used these generative AI services to assist with drafting and refining parts of the manuscript. After using these services, the authors reviewed and edited the content as needed and take full responsibility for the content of the published article.

\FloatBarrier

\bibliographystyle{cas-model2-names}
\bibliography{references}

@article{MDC2021,
    AUTHOR={Mitsufuji, Yuki  and Fabbro, Giorgio  and Uhlich, Stefan  and St{\"o}ter, Fabian-Robert  and Défossez, Alexandre  and Kim, Minseok  and Choi, Woosung  and Yu, Chin-Yun  and Cheuk, Kin-Wai },
    TITLE={Music Demixing Challenge 2021},
    JOURNAL={Frontiers in Signal Processing},
    VOLUME={1},
    YEAR={2022},
    doi={10.3389/frsip.2021.808395},  
    ISSN={2673-8198}
  }

@inproceedings{rouard2023htdemucs,
  author    = {Rouard, Simon and Massa, Francisco and D{\'e}fossez, Alexandre},
  title     = {Hybrid Transformers for Music Source Separation},
  booktitle = {IEEE International Conference on Acoustics, Speech and Signal Processing (ICASSP)},
  year      = {2023},
  doi       = {10.1109/ICASSP49357.2023.10096956},
  note      = {arXiv:2211.08553}
}

@inproceedings{defossez2021hybrid,
  author    = {D{\'e}fossez, Alexandre},
  title     = {Hybrid Spectrogram and Waveform Source Separation},
  booktitle = {Proceedings of the International Society for Music Information Retrieval 2021 Workshop on Music Source Separation (MDX)},
  year      = {2021},
  doi      = {10.48550/arXiv.2111.03600}
}

@article{defossez2019demucs,
  author  = {D{\'e}fossez, Alexandre and Usunier, Nicolas and Bottou, L{\'e}on and Bach, Francis},
  title   = {Music Source Separation in the Waveform Domain},
  url={https://arxiv.org/abs/1911.13254},
  journal = {arXiv preprint arXiv:1911.13254},
  year    = {2019}
}

@misc{defossez_demucs_repo,
  author       = {D{\'e}fossez, Alexandre},
  title        = {Demucs: Code for the paper Hybrid Spectrogram and Waveform Source Separation},
  year         = {2023},
  publisher    = {GitHub},
  howpublished = {\url{https://github.com/adefossez/demucs}},
  note         = {Accessed: 2026-04-13}
}

@inproceedings{stoller2018waveunet,
  author    = {Stoller, Daniel and Ewert, Sebastian and Dixon, Simon},
  title     = {Wave-{U}-{N}et: A Multi-Scale Neural Network for End-to-End Audio Source Separation},
  booktitle = {Proceedings of the 19th International Society for Music Information Retrieval Conference},
  year      = {2018},
  note      = {arXiv:1806.03185}
}

@misc{kim2021kuielabmdxnettwostreamneuralnetwork,
      title={KUIELab-MDX-Net: A Two-Stream Neural Network for Music Demixing}, 
      author={Minseok Kim and Woosung Choi and Jaehwa Chung and Daewon Lee and Soonyoung Jung},
      year={2021},
      eprint={2111.12203},
      archivePrefix={arXiv},
      primaryClass={eess.AS},
      url={https://arxiv.org/abs/2111.12203}, 
}

@article{luo2023bsrnn,
  author  = {Luo, Yi and Yu, Jianwei},
  title   = {Music Source Separation with Band-Split {RNN}},
  journal = {IEEE/ACM Transactions on Audio, Speech, and Language Processing},
  volume  = {31},
  number  = {},
  pages   = {1893--1901},
  year    = {2023},
  doi     = {10.1109/TASLP.2023.3271145}
}

@INPROCEEDINGS{lu2024bsroformer,
  author={Lu, Wei-Tsung and Wang, Ju-Chiang and Kong, Qiuqiang and Hung, Yun-Ning},
  booktitle={ICASSP 2024 - 2024 IEEE International Conference on Acoustics, Speech and Signal Processing (ICASSP)}, 
  title={Music Source Separation With Band-Split Rope Transformer}, 
  year={2024},
  volume={},
  number={},
  pages={481-485},
  doi={10.1109/ICASSP48485.2024.10446843}}

@misc{wang2023melbandroformermusicsource,
      title={Mel-Band RoFormer for Music Source Separation}, 
      author={Ju-Chiang Wang and Wei-Tsung Lu and Minz Won},
      year={2023},
      eprint={2310.01809},
      archivePrefix={arXiv},
      primaryClass={cs.SD},
      url={https://arxiv.org/abs/2310.01809}, 
}

@article{vincent2006bsseval,
  author  = {Vincent, Emmanuel and Gribonval, R{\'e}mi and F{\'e}votte, C{\'e}dric},
  title   = {Performance Measurement in Blind Audio Source Separation},
  journal = {IEEE Transactions on Audio, Speech, and Language Processing},
  volume  = {14},
  number  = {4},
  pages   = {1462--1469},
  year    = {2006},
  doi     = {10.1109/TSA.2005.858005}
}

@misc{MUSDB18HQ,
  author       = {Rafii, Zafar and
                  Liutkus, Antoine and
                  Fabian-Robert St{\"o}ter and
                  Mimilakis, Stylianos Ioannis and
                  Bittner, Rachel},
  title        = {{MUSDB18-HQ} - an uncompressed version of MUSDB18},
  month        = dec,
  year         = 2019,
  doi          = {10.5281/zenodo.3338373},
  url          = {https://doi.org/10.5281/zenodo.3338373}
}

@inproceedings{jansson2017singing,
  title={Singing Voice Separation with Deep U-Net Convolutional Networks},
  author={Jansson, Andreas and Humphrey, Eric and Montecchio, Nicola and Bittner, Rachel and Kumar, Aparna and Weyde, Tillman},
  booktitle={18th international society for music information retrieval conference},
  year={2017}
}

@article{Stoeter2019,
    author = {Stöter, Fabian-Robert and Uhlich, Stefan and Liutkus, Antoine and Mitsufuji, Yuki},
    title = {Open-Unmix - A Reference Implementation for Music Source Separation},
    journal = {Journal of Open Source Software}, 
    doi = {10.21105/joss.01667},
    url = {https://doi.org/10.21105/joss.01667},
    year = {2019},
    publisher = {The Open Journal},
    volume = {4},
    number = {41},
    pages = {1667}
}

@article{xu2013speechenancement,
  author={Xu, Yong and Du, Jun and Dai, Li-Rong and Lee, Chin-Hui},
  journal={IEEE Signal Processing Letters}, 
  title={An Experimental Study on Speech Enhancement Based on Deep Neural Networks}, 
  year={2014},
  volume={21},
  number={1},
  pages={65-68},
  doi={10.1109/LSP.2013.2291240}}

@inproceedings{huang2014singing,
  title={Singing-Voice Separation from Monaural Recordings Using Deep Recurrent Neural Networks},
  author={Huang, Po-Sen and Kim, Minje and Hasegawa-Johnson, Mark and Smaragdis, Paris},
  booktitle={Proceedings of the 15th International Society for Music Information Retrieval Conference},
  pages={477--482},
  year={2014}
}

@inproceedings{UhlichMusicDNN,
  author={Uhlich, Stefan and Porcu, Marcello and Giron, Franck and Enenkl, Michael and Kemp, Thomas and Takahashi, Naoya and Mitsufuji, Yuki},
  booktitle={2017 IEEE International Conference on Acoustics, Speech and Signal Processing (ICASSP)}, 
  title={Improving music source separation based on deep neural networks through data augmentation and network blending}, 
  year={2017},
  volume={},
  number={},
  pages={261-265},
  doi={10.1109/ICASSP.2017.7952158}}

@inproceedings{Narayanan2013maskestimation,
  author={Narayanan, Arun and Wang, DeLiang},
  booktitle={2013 IEEE International Conference on Acoustics, Speech and Signal Processing}, 
  title={Ideal ratio mask estimation using deep neural networks for robust speech recognition}, 
  year={2013},
  volume={},
  number={},
  pages={7092-7096},
  doi={10.1109/ICASSP.2013.6639038}}

@INPROCEEDINGS{Nugraha2016WienerPostDeep,
  author={Nugraha, Aditya Arie and Liutkus, Antoine and Vincent, Emmanuel},
  booktitle={2016 24th European Signal Processing Conference (EUSIPCO)}, 
  title={Multichannel music separation with deep neural networks}, 
  year={2016},
  volume={},
  number={},
  pages={1748-1752},
  doi={10.1109/EUSIPCO.2016.7760548}}

@INPROCEEDINGS{Luo2018TaSnet,
  author={Luo, Yi and Mesgarani, Nima},
  booktitle={2018 IEEE International Conference on Acoustics, Speech and Signal Processing (ICASSP)}, 
  title={TaSNet: Time-Domain Audio Separation Network for Real-Time, Single-Channel Speech Separation}, 
  year={2018},
  volume={},
  number={},
  pages={696-700},
  doi={10.1109/ICASSP.2018.8462116}}

@ARTICLE{luo2019convtasnet,
  author={Luo, Yi and Mesgarani, Nima},
  journal={IEEE/ACM Transactions on Audio, Speech, and Language Processing}, 
  title={Conv-TasNet: Surpassing Ideal Time–Frequency Magnitude Masking for Speech Separation}, 
  year={2019},
  volume={27},
  number={8},
  pages={1256-1266},
  doi={10.1109/TASLP.2019.2915167}}

@article{MDC2023,
   title={The Sound Demixing Challenge 2023 – Music Demixing Track},
   volume={7},
   ISSN={2514-3298},
   doi={10.5334/tismir.171},
   number={1},
   journal={Transactions of the International Society for Music Information Retrieval},
   publisher={Ubiquity Press, Ltd.},
   author={Fabbro, Giorgio and Uhlich, Stefan and Lai, Chieh-Hsin and Choi, Woosung and Martínez-Ramírez, Marco and Liao, Weihsiang and Gadelha, Igor and Ramos, Geraldo and Hsu, Eddie and Rodrigues, Hugo and Stöter, Fabian-Robert and Défossez, Alexandre and Luo, Yi and Yu, Jianwei and Chakraborty, Dipam and Mohanty, Sharada and Solovyev, Roman and Stempkovskiy, Alexander and Habruseva, Tatiana and Goswami, Nabarun and Harada, Tatsuya and Kim, Minseok and Lee, Jun Hyung and Dong, Yuanliang and Zhang, Xinran and Liu, Jiafeng and Mitsufuji, Yuki},
   year={2024},
   pages={63--84}
   }

@inproceedings{kavalerov2019universal,
  author={Kavalerov, Ilya and Wisdom, Scott and Erdogan, Hakan and Patton, Brian and Wilson, Kevin and Le Roux, Jonathan and Hershey, John R.},
  booktitle={2019 IEEE Workshop on Applications of Signal Processing to Audio and Acoustics (WASPAA)}, 
  title={Universal Sound Separation}, 
  year={2019},
  volume={},
  number={},
  pages={175-179},
  doi={10.1109/WASPAA.2019.8937253}}

@article{liu2023audiosep,
  author={Liu, Xubo and Kong, Qiuqiang and Zhao, Yan and Liu, Haohe and Yuan, Yi and Liu, Yuzhuo and Xia, Rui and Wang, Yuxuan and Plumbley, Mark D. and Wang, Wenwu},
  journal={IEEE Transactions on Audio, Speech and Language Processing}, 
  title={Separate Anything You Describe}, 
  year={2025},
  volume={33},
  number={},
  pages={458-471},
  doi={10.1109/TASLP.2024.3520017}}

@INPROCEEDINGS{wu2023clap,
  author={Wu, Yusong and Chen, Ke and Zhang, Tianyu and Hui, Yuchen and Berg-Kirkpatrick, Taylor and Dubnov, Shlomo},
  booktitle={ICASSP 2023 - 2023 IEEE International Conference on Acoustics, Speech and Signal Processing (ICASSP)}, 
  title={Large-Scale Contrastive Language-Audio Pretraining with Feature Fusion and Keyword-to-Caption Augmentation}, 
  year={2023},
  volume={},
  number={},
  pages={1-5},
  doi={10.1109/ICASSP49357.2023.10095969}}

@misc{samaudio2025,
      title={SAM Audio: Segment Anything in Audio}, 
      author={Bowen Shi and Andros Tjandra and John Hoffman and Helin Wang and Yi-Chiao Wu and Luya Gao and Julius Richter and Matt Le and Apoorv Vyas and Sanyuan Chen and Christoph Feichtenhofer and Piotr Dollár and Wei-Ning Hsu and Ann Lee},
      year={2025},
      eprint={2512.18099},
      archivePrefix={arXiv},
      primaryClass={eess.AS},
      url={https://arxiv.org/abs/2512.18099}, 
}

@INPROCEEDINGS{Subakan2018GANsss,
  author={Subakan, Y. Cem and Smaragdis, Paris},
  booktitle={2018 IEEE International Conference on Acoustics, Speech and Signal Processing (ICASSP)}, 
  title={Generative Adversarial Source Separation}, 
  year={2018},
  volume={},
  number={},
  pages={26-30},
  doi={10.1109/ICASSP.2018.8461671}}

@ARTICLE{Zhu2022GenerativeFlow,
  author={Zhu, Ge and Darefsky, Jordan and Jiang, Fei and Selitskiy, Anton and Duan, Zhiyao},
  journal={IEEE Signal Processing Letters}, 
  title={Music Source Separation With Generative Flow}, 
  year={2022},
  volume={29},
  number={},
  pages={2288-2292},
  doi={10.1109/LSP.2022.3219355}}

@misc{mariani2024multisourcediffusionmodelssimultaneous,
      title={Multi-Source Diffusion Models for Simultaneous Music Generation and Separation}, 
      author={Giorgio Mariani and Irene Tallini and Emilian Postolache and Michele Mancusi and Luca Cosmo and Emanuele Rodolà},
      year={2024},
      eprint={2302.02257},
      archivePrefix={arXiv},
      primaryClass={cs.SD},
      url={https://arxiv.org/abs/2302.02257}, 
}

@INPROCEEDINGS{Scheibler2025flowmatching,
  author={Scheibler, Robin and Hershey, John R. and Doucet, Arnaud and Li, Henry},
  booktitle={2025 IEEE Workshop on Applications of Signal Processing to Audio and Acoustics (WASPAA)}, 
  title={Source Separation by Flow Matching}, 
  year={2025},
  volume={},
  number={},
  pages={1-5},
  doi={10.1109/WASPAA66052.2025.11230963}}

@ARTICLE{Richter2023diffusionSpeechEnh,
  author={Richter, Julius and Welker, Simon and Lemercier, Jean-Marie and Lay, Bunlong and Gerkmann, Timo},
  journal={IEEE/ACM Transactions on Audio, Speech, and Language Processing}, 
  title={Speech Enhancement and Dereverberation With Diffusion-Based Generative Models}, 
  year={2023},
  volume={31},
  number={},
  pages={2351-2364},
  doi={10.1109/TASLP.2023.3285241}}

@INPROCEEDINGS{Yuan2025flowsep,
  author={Yuan, Yi and Liu, Xubo and Liu, Haohe and Plumbley, Mark D. and Wang, Wenwu},
  booktitle={ICASSP 2025 - 2025 IEEE International Conference on Acoustics, Speech and Signal Processing (ICASSP)}, 
  title={FlowSep: Language-Queried Sound Separation with Rectified Flow Matching}, 
  year={2025},
  volume={},
  number={},
  pages={1-5},
  doi={10.1109/ICASSP49660.2025.10890129}}

@INPROCEEDINGS{Wang2025ExtractDiffTrans,
  author={Wang, Helin and Hai, Jiarui and Lu, Yen-Ju and Thakkar, Karan and Elhilali, Mounya and Dehak, Najim},
  booktitle={ICASSP 2025 - 2025 IEEE International Conference on Acoustics, Speech and Signal Processing (ICASSP)}, 
  title={SoloAudio: Target Sound Extraction with Language-oriented Audio Diffusion Transformer}, 
  year={2025},
  volume={},
  number={},
  pages={1-5},
  doi={10.1109/ICASSP49660.2025.10890066}}

@ARTICLE{8528831,
  author={Ar{\i}k, Sercan {\"O}. and Jun, Heewoo and Diamos, Gregory},
  journal={IEEE Signal Processing Letters}, 
  title={Fast Spectrogram Inversion Using Multi-Head Convolutional Neural Networks}, 
  year={2019},
  volume={26},
  number={1},
  pages={94-98},
  doi={10.1109/LSP.2018.2880284}}

@inproceedings{perez2018film,
  author    = {Perez, Ethan and Strub, Florian and de Vries, Harm and Dumoulin, Vincent and Courville, Aaron},
  title     = {{FiLM}: Visual Reasoning with a General Conditioning Layer},
  booktitle = {Proceedings of the AAAI Conference on Artificial Intelligence},
  year      = {2018},                                                                                                             
  doi={10.1609/aaai.v32i1.11671}
}

@inproceedings{meseguer2019cunet,
  author    = {Meseguer-Brocal, Gabriel and Peeters, Geoffroy},
  title     = {Conditioned-{U}-{N}et: Introducing a Control Mechanism in the U-Net for Multiple Source Separations},
  booktitle = {Proceedings of the 20th International Society for Music Information Retrieval Conference},
  pages     = {159--165},
  year      = {2019},
  note      = {arXiv:1907.01277}
}

@inproceedings{choi2021conditioning-instruments,
  author={Choi, Woosung and Kim, Minseok and Chung, Jaehwa and Jung, Soonyoung},
  booktitle={ICASSP 2021 - 2021 IEEE International Conference on Acoustics, Speech and Signal Processing (ICASSP)}, 
  title={{LaSAFT}: Latent Source Attentive Frequency Transformation For Conditioned Source Separation}, 
  year={2021},
  volume={},
  number={},
  pages={171--175},
  doi={10.1109/ICASSP39728.2021.9413896}}

@misc{tunturi2025scoreinformedmusicsourceseparation,
      title={Score-informed Music Source Separation: Improving Synthetic-to-real Generalization in Classical Music}, 
      author={Eetu Tunturi and David Diaz-Guerra and Archontis Politis and Tuomas Virtanen},
      year={2025},
      eprint={2503.07352},
      archivePrefix={arXiv},
      primaryClass={eess.AS},
      url={https://arxiv.org/abs/2503.07352}, 
}

@misc{jeon2020exploringalignedlyricsinformedsinging,
      title={Exploring Aligned Lyrics-Informed Singing Voice Separation}, 
      author={Chang-Bin Jeon and Hyeong-Seok Choi and Kyogu Lee},
      year={2020},
      eprint={2008.04482},
      archivePrefix={arXiv},
      primaryClass={eess.AS},
      url={https://arxiv.org/abs/2008.04482}, 
}

@inproceedings{gover2020score,
    author = {Gover, Matan and Depalle, Philippe},
    title = {Score-informed source separation of choral music},
    booktitle = {Proceedings of the 21st International Society for Music Information Retrieval Conference},
    pages = {231--239},
    year = {2020}
}

@inproceedings{miron2017monaural,
  title={Monaural score-informed source separation for classical music using convolutional neural networks},
  author={Miron, Marius and Janer Mestres, Jordi and G{\'o}mez, Emilia},
  booktitle={Proceedings of the 18th International Society for Music Information Retrieval Conference},
  year={2017},
  pages={55--62}}

@article{gulli2025enhancing,
  title={Enhancing drone audition with rotor-conditioned deep models},
  author={Gulli, Andrea and Fontana, Federico and Drioli, Carlo and Salvati, Daniele and Ferrin, Giovanni},
  journal={EURASIP Journal on Audio, Speech, and Music Processing},
  volume={2025},
  number={1},
  pages={35},
  year={2025},
  doi={10.1186/s13636-025-00425-2}
}

@inproceedings{tobin2017domainrand,
  author={Tobin, Josh and Fong, Rachel and Ray, Alex and Schneider, Jonas and Zaremba, Wojciech and Abbeel, Pieter},
  booktitle={2017 IEEE/RSJ International Conference on Intelligent Robots and Systems (IROS)}, 
  title={Domain randomization for transferring deep neural networks from simulation to the real world}, 
  year={2017},
  volume={},
  number={},
  pages={23-30},
  doi={10.1109/IROS.2017.8202133}}

@article{froehlingsdorf2023atz,
  author  = {Fr{\"o}hlingsdorf, Katja and Doleschal, Florian and Pischinger, Stefan and Verhey, Jesko L.},
  title   = {Efficient Analysis of the Interior Noise of Electric Vehicles},
  journal = {ATZ worldwide},
  volume  = {125},
  number  = {1},
  pages   = {58--63},
  year    = {2023},
  doi     = {10.1007/s38311-022-1429-z}
}

@article{froehlingsdorf2023imeche,
  author  = {Fr{\"o}hlingsdorf, Katja and Pischinger, Stefan and G{\"u}nther, Marco and Steffens, Christoph and Heuer, Stephan},
  title   = {Automated Inverter Noise Separation in the Interior of Electric Vehicles Using Adapted Kirsch-Compass Kernel and Hough-Transformation},
  journal = {Proceedings of the Institution of Mechanical Engineers, Part D: Journal of Automobile Engineering},
  volume  = {237},
  number  = {2--3},
  year    = {2023},
  pages={312--320},
  doi     = {10.1177/09544070221076591}
}

@phdthesis{froehlingsdorf2023diss,
  author = {Fr{\"o}hlingsdorf, Katja},
  title  = {Automatisierte Trennung und Zuordnung von Ger{\"a}uschanteilen im Innenraum von Elektrofahrzeugen},
  school = {RWTH Aachen University},
  year   = {2023},
  doi    = {10.18154/RWTH-2023-08448}
}

@INPROCEEDINGS{fvv1369,
      author       = {Pischinger, Stefan and Verhey, Jesko and Fröhlingsdorf,
                      Katja and Doleschal, Florian},
      title        = {{S}törgeräusche im {F}ahrzeuginnenraum mit
                      elektrifizierten {A}ntrieben - {K}enngrößen zur
                      {Q}uantifizierung der {L}ästigkeit von {S}törgeräuschen
                      im {I}nnenraum von {F}ahrzeugen mit elektrifiziertem
                      {A}ntrieb},
      publisher    = {FVV},
      series       = {Forschungsvereinigung Verbrennungskraftmaschinen. R ; 603},
      pages        = {494--528},
      year         = {2022},
      booktitle     = {The FVV Transfer+Networking Event,
                       October 2022 : Final and interim
                       reports presented by the RTD performers
                       : proceedings (2022)
                       Forschungsvereinigung
                       Verbrennungskraftmaschinen 2022
                       Würzburg Herbsttagung},
      cin          = {412310},
      cid          = {$I:(DE-82)412310_20140620$},
      typ          = {PUB:(DE-HGF)7 / PUB:(DE-HGF)8},
      url          = {https://publications.rwth-aachen.de/record/856638},
}

@inproceedings{bugl2025daga,
  author    = {B. Bugl and H.-P. Rabl and T{\"o}pken, Stephan and E. Claaßen and S. van de Par and A. Rieger},
  title     = {Implementation of a synthesis tool for stationary and non-stationary interior noise of electric vehicles},
  booktitle = {Proceedings of DAS|DAGA 2025},
  year      = {2025},
  pages   = {1499--1502}
}

@article{jiang2024fldnet,
  author  = {Jiang, DingLin and Luo, Xinwei and Shen, Qifan},
  title   = {Frequency Line Detection in Spectrograms Using a Deep Neural Network with Attention},
  journal = {The Journal of the Acoustical Society of America},
  volume  = {156},
  number  = {5},
  pages   = {3204--3216},
  year    = {2024},
  doi     = {10.1121/10.0034360}
}

@misc{prakash2022bayesiannvhmetamodelsassess,
      title={Bayesian NVH metamodels to assess interior cabin noise using measurement databases}, 
      author={V. Prakash and O. Sauvage and J. Antoni and L. Gagliardini},
      year={2022},
      eprint={2207.02120},
      archivePrefix={arXiv},
      primaryClass={stat.AP},
      url={https://arxiv.org/abs/2207.02120}, 
}

@article{prakash2025stochastic,
  title={Stochastic assessment of electric powertrain whining noise under early-stage design uncertainties},
  author={Prakash, Vinay and Sauvage, Olivier and Antoni, J{\'e}r{\^o}me and Gagliardini, Laurent and Totaro, Nicolas},
  journal={Proceedings of the Institution of Mechanical Engineers, Part D: Journal of Automobile Engineering},
  volume={239},
  number={8},
  pages={3246--3265},
  year={2025},
  publisher={SAGE Publications Sage UK: London, England}
}

@article{LI2018tirereview,
title = {Literature review of models on tire-pavement interaction noise},
journal = {Journal of Sound and Vibration},
volume = {420},
pages = {357--445},
year = {2018},
issn = {0022-460X},
doi = {10.1016/j.jsv.2018.01.026},
url = {https://www.sciencedirect.com/science/article/pii/S0022460X18300348},
author = {Tan Li and Ricardo Burdisso and Corina Sandu},
}

@article{oboy2009tiremodel,
title = {Tyre/road interaction noise—A 3D viscoelastic multilayer model of a tyre belt},
journal = {Journal of Sound and Vibration},
volume = {322},
number = {4--5},
pages = {829--850},
year = {2009},
issn = {0022-460X},
doi = {10.1016/j.jsv.2008.09.021},
url = {https://www.sciencedirect.com/science/article/pii/S0022460X0800802X},
author = {D.J. O’Boy and A.P. Dowling},
}

@inproceedings{koners2009airbornetire,
  author = "Koners, Gregor and Lehmann, Ralf",
  title = "Investigation of Tire Road Noise with Special Consideration of Airborne Noise Transmission",
  booktitle = "SAE 2009 Noise and Vibration Conference and Exhibition",
  publisher = "SAE International",
  month = may,
  year = 2009,
  doi = "10.4271/2009-01-2109",
  url = "https://doi.org/10.4271/2009-01-2109",
  issn = "0148-7191",
}

@article{bederna2009contributions,
  title={Contributions to a better understanding of tire cavity noise},
  author={Bederna, Ch. and Saemann, E.-U.},
  journal={Deutsche Gesellschaft f{\"u}r Akustik eV (DEGA): DAGA-Jahrestagung f{\"u}r Akustik},
  volume={2009},
  pages={502--505},
  year={2009}
}

@article{cerrato2009windnoise,
  title={Automotive Sound Quality – Powertrain, Road and Wind Noise},
  author={Cerrato, Gabriella},
  journal={Sound and Vibration},
  volume={43},
  number={4},
  pages={16--24},
  year={2009}
}

@article{oettle2017automotiveaeroacoustics,
  title={Automotive aeroacoustics: An Overview},
  author={Oettle, Nicholas and Sims-Williams, David},
  journal={Proceedings of the Institution of Mechanical Engineers, Part D: Journal of Automobile Engineering},
  volume={231},
  number={9},
  pages={1177--1189},
  year={2017},
  publisher={SAGE Publications Sage UK: London, England}
}

@incollection{blumrich2006computational,
  title={Computational Aeroacoustics in Automotive Engineering},
  author={Blumrich, R.},
  booktitle={Progress in Vehicle Aerodynamics IV: Numerical Methods}, 
  publisher={Expert Verlag}, 
  year={2006}
}

@inproceedings{Wickern2009Aeroacoustics,
  author = "Wickern, Gerhard and Brennberger, Martin",
  title = "Scaling Laws in Automotive Aeroacoustics",
  booktitle = "SAE World Congress \& Exhibition",
  publisher = "SAE International",
  month = apr,
  year = 2009,
  doi = "10.4271/2009-01-0180",
  url = "https://doi.org/10.4271/2009-01-0180",
  issn = "0148-7191"
}

@ARTICLE{Vijayraghavan1999emNoiseReview,
  author={Vijayraghavan, P. and Krishnan, R.},
  journal={IEEE Transactions on Industry Applications}, 
  title={Noise in electric machines: a review}, 
  year={1999},
  volume={35},
  number={5},
  pages={1007-1013},
  doi={10.1109/28.793360}}

@inproceedings{kang2019electromagneticNoiseEM,
  title={Test and Analysis of Electromagnetic Noise of an Electric Motor in a Pure Electric Car},
  author={Kang, Qiang and Gu, Perry and Gong, Chao and Zuo, Shuguang},
  booktitle={Noise and Vibration Conference \& Exhibition},
  year={2019},
  doi= {10.4271/2019-01-1492}
}

@inproceedings{mehrgou2018EMgearboxinv,
  title={{NVH} Aspects of Electric Drives-Integration of Electric Machine, Gearbox and Inverter},
  author={Mehrgou, Mehdi and Garcia de Madinabeitia, Inigo and Graf, Bernhard and Zieher, Franz and Priestner, Christoph},
  booktitle={10th International Styrian Noise, Vibration \& Harshness Congress: The European Automotive Noise Conference},
  year={2018}
}

@inproceedings{patra2026HVAC,
  title={Acoustic Challenges in Electric Vehicles: A Comparative Study of HVAC Noise in EVs and ICE Vehicles},
  author={Patra, Subhashree and Joshi, Rishi and Sharma, Rachit and Lingampelly, Rajaprasad and Neupane, Manoj},
  booktitle={Symposium on International Automotive Technology (2026)},
  year={2026},
  doi={10.4271/2026-26-0353}
}

@inproceedings{titave2025designHVAC,
  title={Design Strategies for Integrating a Quieter Refrigeration Handling System in Automotive HVAC},
  author={Titave, Uttam Vasant and Kalsule, Shrikant and Naidu, Sudhakara},
  booktitle={SAENIS TTTMS Thermal Management Systems Conference},
  year={2025},
  doi={10.4271/2025-28-0432}
}

@inproceedings{noumura2006otpa,
  title={Method of Transfer Path Analysis for Vehicle Interior Sound with No Excitation Experiment},
  author={Noumura, Kousuke and Yoshida, Junji},
  year={2006},
  booktitle={FISITA 2006 World Automotive Congress}
}

@article{gajdatsy2008critical,
  title={Critical assessment of Operational Path Analysis: mathematical problems of transmissibility estimation},
  author={Gajdatsy, Peter and Janssens, Karl and Gielen, Ludo and Mas, Peter and Van Der Auweraer, Herman},
  journal={Journal of the Acoustical Society of America},
  volume={123},
  number={5},
  pages={3869},
  year={2008},
  doi={10.1121/1.2935749}
}

@article{tpa2016framework,
  author  = {van der Seijs, Maarten V. and de Klerk, Dennis and Rixen, Daniel J.},
  title   = {General framework for transfer path analysis: History, theory and classification of techniques},
  journal = {Mechanical Systems and Signal Processing},
  volume  = {68--69},
  pages   = {217--244},
  year    = {2016},
  doi     = {10.1016/j.ymssp.2015.08.004}
}

\section{Appendix}
\begin{table}[htbp]
\centering
\caption{Randomly sampled properties, their value ranges and underlying distributions.}
\label{tab:random_property_distributions}
\small
\begin{tabular}{p{4.1cm} p{2.0cm} p{4.4cm} p{2.7cm}}
\toprule
\textbf{Property} & \textbf{Scope} & \textbf{Value Range} & \textbf{Distribution} \\
\midrule

Tire exterior diameter [m] & per vehicle & $[0.60,\,0.75]$ & Uniform \\
Gear ratio & per vehicle & $[7.5,\,11.5]$ & Uniform \\
Pole pairs $p$ & per vehicle & $\{3,4,5\}$ & Discrete uniform \\
Stator slots $N_{\text{slots}}$& per vehicle & $\{36,48,72\}$ & Discrete uniform \\
Gear teeth count $N_{\text{teeth}}$ & per vehicle & $[80, 100]$ & Uniform, integer \\
Inverter switching freq.\ $f_{sw}$ [Hz] & per vehicle & $[7000,\,14000]$ & Uniform \\
Cavity mode base freq. at 60 km/h\ [Hz] & per vehicle & $[170,\,230]$ & Uniform \\
Cavity mode amplitude scaling & per vehicle & $\mu$ = 1.0, $\sigma=0.025$ & Normal \\
offset for ratio of broadband noise (tire + wind) to tonal powertrain noise [dB] & per vehicle & $[-0.5,\,0.5]$ & Uniform \\
Wind amplification factor & per vehicle & $[0.9,\,1.1]$ & Uniform \\
\# active drive unit orders (mechanical) & per vehicle & 3--11 (drawn from pool of orders $\{2,\dots,12\}$, order 1 always included) & Discrete uniform, integer \\
\# active drive unit orders (electrical) & per vehicle & 3--5 (drawn from pool $\{2,\dots,6\}$, order 1 always included) & Discrete uniform, integer \\
Order amplitude scaling & per vehicle, per order & clipped to $[0.5,\,2.0]$, $\mu$ = 1.0, $\sigma=0.2$ & Normal (clipped) \\
Order amplitude scaling & per order, per sample & $[0.95,\,1.05]$ & Uniform \\
Inverter noise amplitude scaling (pt\_inv mixing) & per vehicle & $[0.03,\,0.08]$ & Uniform \\
Auxiliary drive amplitude scaling & per vehicle & $[0.003,\,0.008]$ & Uniform \\
Cabin Transfer Function (TF): \# resonance peaks & per TF instance & 3--7 & Discrete uniform \\
Cabin TF: resonance frequencies shift & per resonance peak & log-spaced resonance frequencies \ (200--5000\,Hz range) shifted with a factor $[0.8,\,1.2]$  & Uniform \\
Cabin TF: resonance bandwidth & per resonance peak & $[0.2,\,0.5]$ & Uniform \\
Cabin TF: resonance gain & per resonance peak & $[1.2,\,2.0]$ & Uniform \\
Separation-IR level variation & per vehicle & small \ frequency-domain randomization applied, clipped to $\pm 10$\,dB & Normal (smoothed, clipped) \\
Aux base rotational speed & per sample & $[900,\,3500]$ & Uniform \\
Aux rotational speed drift & per sample & smoothed random walk, typical peak-to-peak $\approx$150 RPM over 10\,s & Cumulative Gaussian random walk, 2nd-order low-pass (0.5\,Hz) \\
Driving segment & per sample & 10 s duration and of 3 categories: accel/cruise/decel; speed range $0 - 130 km/h$; max. $\Delta v = 60 km/h$  & Categorical uniform \\

\bottomrule
\end{tabular}
\end{table}

\begin{table}[htbp]
\centering
\caption{Influence of vehicle operation parameters on the synthesized sound sources.}
\label{tab:operation_parameter_influence}
\small
\begin{tabular}{p{3.1cm} p{2.6cm} p{8.3cm}}
\toprule
\textbf{Operation Parameter} & \textbf{Affected Source(s)} & \textbf{Mechanism / Effect} \\
\midrule

Vehicle speed $v(t)$ [km/h]
& Powertrain, Inverter
& Sets drive unit rotational speed (see below) \\

& Tire, Wind (broadband)
& Broadband noise (sum of tire and wind) amplitude scales linearly with vehicle speed \citep{bugl2025daga} \\

& Wind only
& Low-speed roll-off via sigmoid $1/(1+e^{-0.2(v-20)})$ gradually suppresses wind noise below $\sim$50 km/h \\

& Cavity tone (tire sub-component)
& Resonance frequency shifts with speed: $f_{\text{cav}} = f_{\text{base}} + 0.1\,(v-60 $ km/h$)$; amplitude scales relative to tire RMS \\

\midrule
Load factor $L(t)\in[0,1]$ (derived)
& Powertrain
& Time-varying, constructed with speed-dependence and acceleration-dependence. Modulates amplitude of all powertrain orders.
\\

\midrule
Drive unit rotational speed $f_{\text{mech}}$ (derived from $v$ via gear ratio and wheel diameter)
& Powertrain, Inverter
& Determines exact excitation frequencies: mechanical orders $f_{\text{m}}$ = $n_m\cdot f_{\text{mech}}$ ; electrical orders $f_{\text{e}}$ = $n_e\cdot f_{\text{mech}}\cdot p$ ($p$ = pole pairs); slot order $f_{\text{slot}}$ =$f_{\text{mech}}\cdot N_{\text{slots}}$; gear mesh order $f_{\text{gear}}$ = $f_{\text{mech}}\cdot N_{\text{teeth}}$; inverter sidebands $f_{\text{sb}}$ = $f_{sw}\pm j\cdot f_{\text{mech}}\cdot p$ (sidebands $j=1,\dots,N_{sb}$); all order amplitudes increase with higher rotational speeds \\

\midrule
Auxiliary drive rotational speed (random, independent from driving scenario)
& Aux only
& Determines exact excitation frequencies of auxiliary drive orders\\

\bottomrule
\end{tabular}
\end{table}

\end{document}